# User-driven applications

***Abstract.*** User-driven applications are the programs, in which the full control is given to the users. Designers of such programs are responsible only for developing an instrument for solving some task, but they do not enforce users to work with this instrument according with the predefined scenario. Users' control of the applications means that only users decide at any moment WHAT, WHEN, and HOW must appear on the screen. Such applications can be constructed only on the basis of moveable / resizable elements. Programs, based on such elements, have very interesting features and open absolutely new possibilities. This article describes the design of the user-driven applications and shows the consequences of switching to such type of programs on the samples from different areas.

## Content

## Introduction

This is the second article in the series of two. The first article – ***"On the theory of moveable objects"*** – was about the basic algorithm and its use in design of different moveable / resizable objects. Such objects are at high demand in many programming areas, so the algorithm can be widely used by itself. But from my point of view the consequences of using the moveable objects as the basis of design are much more important than the algorithm itself. Regardless of the algorithm's origin, if you start to design your programs on the basis of moveable / resizable objects, you'll come to the same situation: you'll find yourself in another programming world. The design of moveable elements and the consequences of using such elements belong to two different, but strongly related theories; that's why I decided to discuss the whole theme in two separate articles.

There are less code samples in the text of this article, but more explanations, than in the first one. The main idea of this article is to show the use of total moveability for mass production of applications. (Mass production, thought out and organized by Henry Ford, is much more important for the auto manufacturing, than the design of some new detail or even a car.) Less code samples in this text doesn't mean that this article is less supported with the samples. The accompanying application consists of several samples; some of them were developed and used as stand alone applications and only later were included here. For design of moveable objects, the code details are important, so the first article was concentrated on them. For development of user-driven applications, the simultaneous implementation of several rules is important. All the pieces of code are available in the samples; but the explanation to these pieces and their overall behaviour become the most important thing.

Moveable objects can be used in applications from absolutely different areas. These objects can be of different shapes and purposes; they can differ so much that it would be difficult to find any common features between them, except the moveability. And yet there is an algorithm, which allows to turn any object into moveable / resizable. Programmers are the people, who are responsible for design of the screen objects with one or another set of features, so the ***"Theory of moveable objects"*** was written especially for programmers as DEVELOPERS. But applications on moveable / resizable elements – *user-driven applications* – give absolutely new possibilities to the users. All of us, programmers and designers, use for our work the programs, developed by somebody else, so for majority of our work time we are users. This article is written for programmers AS USERS, because only after understanding the difference between the currently used programs and the described user-driven applications, each programmer can decide for himself, if he agrees with the importance of switching to another type of applications, and if he wants to develop his applications in the new way. You have to decide for yourself, if you want those programs that you use all the time, to be totally controlled by you, or are you satisfied with the current situation, when you work with any program under the strict limitations, enforced by their designers. It's your personal choice; nobody else can decide it for you. If you are going to continue with this article, then try to estimate each statement not only professionally as a programmer, but also evaluate it AS A USER of the programs, in which those ideas can be implemented.

Applications, based on the moveable / resizable elements, differ so much from what we got used to for many years, that it is difficult to accept them at first. This is the nature of all the human beings: if we see the thing, which differs 10-15 percent from the usual things, we can estimate it without any bias. But if we are introduced to something that is nearly 100 percent different from the familiar things, then we have big problems not only in understanding it, but we prefer not even to try it. Keep it in mind, when you'll look at the accompanying program, and without looking at it and trying it in different ways you'll not understand the new ideas.

I hope that you know, how difficult it was for people to accept the idea that not the Sun was going around the Earth, but vise worth. It was especially difficult, because the idea that the Earth was the center of the Universe wasn't one of the concurrent theories, which at one moment turned out to be wrong. It was an AXIOM, and axioms are usually not discussed or doubted; they are simply accepted without any second thought.

In programming we have such type of axiom, which was never doubted since the beginning of the programming era: <u>any program is used according with the developer's understanding of the task and the scenario that was coded at the stage of the design</u>. After reading this, you'll definitely announce a statement: "Certainly. How else can it be?" Well, the Sun was going around the Earth for centuries. There were no doubts about it. Yet, it turned out to be not correct.

All programs that we have now, work strictly according with their designers' scenario. (My programs, including the one, which accompanies this article, do not obey that axiom, but the amount of such programs is negligible.) When you develop any program for your personal use, then the design under the underlined axiom is not a problem: if you need something different from the program, you change it. The problem starts, when the number of users starts growing. People have different ideas and preferences; they have different demands.

Computers were invented as pure calculators. Powerful, then more powerful and even more, but for many years they were only pure calculators and nothing else. When the main and the only goal of a program is to give out the results of some calculations, then the only important thing is the correctness of results. The results depend on the initial equations and their coding, so there is nothing wrong if such program works strictly according with the developers' ideas. As this was the only type of programs for decades, then the main rule of their design became an axiom.

The invention of graphical display significantly improved the process of data and results analysis (people are much better in understanding the graph than the column of numbers), but didn't change the main idea of the programs. Then the era of new programs started; the problems began to grow like an avalanche. Even in the specialized scientific / engineering programs the part of calculations was shrinking very quickly. The visualization became the dominant part even in such applications; there are also a lot of programs without any calculations at all, but with the requirements for a wide variety of data and results presentation and for different types of man – machine interfaces. The questions of interface became the most important in the programs' design. When an application is used by hundreds, thousands or millions of users, then it's impossible to imagine that all of them would agree with the designers' view and decisions. Yet the new programs are still developed on the basis of the same axiom, which is absolutely inappropriate for them.

The interface of our programs has hardly changed throughout the last 15 – 18 years. The most popular programming languages change, the environment for the programmers' work is absolutely different, but the results look the same. Only by the view of some controls and several other tiny details the specialist can distinguish the 1994 application from the 2009 program. But these changes are simply seasonal adjustments, nothing else. Why are there no real changes for so many years? I think that to uncover the roots of the problem you'll have to look 20 years back, because the programming philosophy, under which the interface design was going all these years, was incorporated at that time.. Somewhere 20 years ago in one of the hot discussions on the interface design, its future changes and the vector of development, I heard such a declaration of the most popular view: "The clients would have to like whatever we'll give them". Rude expression? Maybe, but only from the point of political correctness. If you skip from all the manuals the standard phrases about the priority of the clients' interests (it's only a mantra, which has to be declared; nothing else) and look at what those clients really get, then you'll immediately see that it is an exact declaration of how a lot (a majority) of designers continue to think.[*] I hear too often that the good specialists in design (they call themselves experts) know much better than any user, how the program must look and work, what kind of interface must be organized.

---

[*] The history of different branches of science and engineering starts at different moments and go on with different speeds, but looks like all of them make the same steps in their evolution. Compare that statement about the programs with the famous declaration from Henry Ford: "Any customer can have a car painted any color that he wants so long as it is black".

There is an obvious and very funny flaw in such point of view. The experts in design declare that because of their huge experience they know better than any user, what he really needs. They look at the users of their programs as the crowd that must be directed and instructed up to the tiny details; from such point of view the users are too dull to decide for themselves the questions of organizing the screen view of the applications. But the same experts in design spend a lot (a bigger part) of their work time as the users of other applications; by the developers of these applications, those users-"experts" are considered as dull as everyone else. Don't you see the funny side of it?

How many times were you really mad even with the most popular applications, because in one or another situation the program was not going on according with your expectations but did something different? I am sure that the huge efforts were made to design those applications in the best way, so that you would be always satisfied with their work. And still you are mad with their response on your commands from time to time. The developers are bad? No, they are excellent. You wanted to do something stupid? I doubt. Both sides are correct, but they have different views, preferences, experience, and habits. It's an absolutely wrong idea to declare that only one view on design is correct and to force the opponent to change his view and to agree with your vision of the best design (in such way the developers always have the upper hand).

To solve the problem of the single interface variant enforced on users with different views, the adaptive interface was introduced. Did it solve the problems? No, only soften them a bit. The popular dynamic layout didn't change anything at all except for some extremely simple situations. If you have a resizable form with a single control inside, then with the resizing of the form this control changes its size so as to fill the same percent of the form's area or keep the constant spaces between the control and the form's borders (classical anchoring). This is very easy to implement and the results will satisfy nearly everyone, though I mentioned two possibilities even for this extremely primitive case. But what about two controls inside? They can resize both, or only the first one, or only the second. And the programmer has two choices: either to put into code the variant he personally prefers (or the marketing department of his company declares to be the best), or to add an instrument, which will allow users to select one of three cases. For the case of only two controls you get three possible variants. What if you have not two, but three, four or more controls inside? Do you think that you would be capable to organize a selection between all the possible variants and would like to introduce users with such a system? I doubt. The only solution that is left for you is to put some variant into the code and explain to the users that this is the best variant, which they would like. That's how all the popular (and less popular) programs are designed. Does it differ from that declaration, which was announced 20 years ago?

What is common for any form of adaptive interface (it can be called dynamic layout or by any other name) that somewhere inside the program there is a model of the users' behaviour. This model is either fixed at the moment of design, or can be refined throughout the work of an application, which is slightly better. The most important thing is that this model was put inside by the designer; after it an application tries to work according with this model as best, as it can. The adaptive interface always looked at as a friendly interface. It can be so, if you absolutely share the designer's view on the interface, but it is a rare situation. More often than not the program becomes obtrusively friendly, or even overwhelmingly obtrusive. Millions of people around the world are getting mad with the applications that are changing in the way they don't want them to change. But the users have no choices: the model of their behaviour and <u>what was thought to be the best</u> for them was already fixed.

By the main idea of their design, all applications can be divided into two groups.

- If the applications are the instruments to solve more and more sophisticated tasks, then being their designer, you will work on the improvement of the current instruments and the design of the new instruments, without which the new tasks can't be solved.
- If the programs are simply some kind of toys to amuse the public and push her into buying new toys every year, then the best policy is to add a couple of colorful strips and buttons from time to time and by means of very aggressive marketing declare that this is a huge step of evolution. (While in reality there is absolutely nothing new.) Unfortunately, that's the way the programming began to move years ago.

What is the difference between the toys and instruments in the world of applications? If you can do only whatever was predetermined by a designer and fixed in code, then it is a toy; if a designer thought about possibilities and provided them without restrictions, then it is an instrument. The toys can be very interesting and sophisticated; they can allow you to do a lot of things; I like the good toys. But there are a lot of things that can't be done with any kind of toys, but only with the instruments. So what is the main difference between the programming toys and instruments? The control over the behaviour of the inner elements and the scenario of the whole work. To turn the applications from being toys into instruments, you have to give users the full control of all the screen elements, thus giving them the full control of an application.

Was there any moment throughout the history of programming, when users get much more control over the screen objects then before? Yes, it happened once, when the multi-window operational systems introduced the idea of moveable windows at the upper level. It was a huge improvement in simultaneous dealing with several programs, but that was the first and the last step in this direction. Do you understand that it has happened 20 and plus years ago and from that time there was nothing else? The proposed switch from currently used applications to user-driven applications means bigger step for users than that old step from DOS to Windows.

We move the real objects in our real life all the time, because we need them to be positioned slightly different at different moments. Our decisions about the relocation and placement of the real objects depend on many different things, but whenever we think that another place would be better, we move the thing (if it is not too heavy) without thinking an extra second. Just move it and that's all; if we would not like it, we will move the thing again. Absolutely the same thing would happen to the screen objects: if they would be easily moveable, we would move them around the screen very often and place them at the most suitable position, as we estimate it at each moment; I see this happening all the time with the user-driven applications.

Certainly, some people understood the need of moveable (by users!) screen objects years ago. Because of the importance of this task, the moveable objects appeared from time to time in different programs. The main problem was that it happened occasionally, when some clever programmers designed such an algorithm for a particular class of objects. I think that because of their complexity, those algorithms were never applied to other objects. (I was also doing such things years ago, and my old algorithm was too complicated to be used in other situations.) There were no algorithms that could be easily used for different types of objects, and there were no attempts to design applications totally on the basis of moveable objects. Those moveable objects in the older applications were only good solutions for particular cases. I want to emphasize and underline again two things:

- There were no algorithms that allowed to turn into moveable / resizable the objects of an arbitrary shape and origin.
- There were no systems or applications totally designed on moveable / resizable elements.

Only an algorithm that can be easily applied to any object, will allow to design the applications totally on moveable / resizable elements. Before turning your attention to the real user-driven applications, I'll write several words about the algorithm that allowed me to turn any object into moveable / resizable. The first article was about this algorithm; the next several phrases are only to remind some terms, because they are used occasionally in the further text.

To make an object moveable / resizable, it receives a ***cover***, consisting of the invisible ***nodes***; each node is responsible either for moving an object or its resizing (reconfiguring). It turned out that it is enough to have three types of nodes (circles, strips, and convex polygons) to transform an arbitrary object into moveable / resizable. There are no restrictions on the number of nodes, their sizes or placement. The object's area can consist of a single piece of an arbitrary shape or it might consist of several not connected parts; it really doesn't matter. The graphical objects are usually moved by any inner point and resized by their border points. The controls are moved and resized by their borders. The sets of objects can be united into groups, which can be moved and resized in different ways. One approach is in moving / resizing a group by its frame; the sizes and positions of the inner elements are then calculated according with some rules, for example, of dynamic layout. Another approach is based on an opposite idea: the inner elements are moveable / resizable individually; the frame is adjustable to the unified area of the inner elements; the group can be moved by the frame or any inner point.

I organized the whole process of moving / resizing exclusively by the mouse; only three mouse events – `MouseDown`, `MouseUp`, and `MouseMove` – are used for all the purposes. Neither these events nor the use of different buttons for different purposes (I prefer to start the forward moving and resizing by the left button; the rotation – by the right one) are fixed anywhere in the code of an algorithm. These are only my suggestions.

I call the programs, based on moveable / resizable objects, ***user-driven applications***. When you get a car, you get an instrument of transportation. Its manual contains some suggestions on maintenance, but there is no fixed list of destinations for this car. You are the driver, you decide about the place to go and the way to go. That's the meaning of the term *user-driven application*: you run the program and make all the decisions about its use; a designer only provides you with an instrument, which allows you to drive.

This article comes with an accompanying program. Only reading the article is definitely not enough; you'll not understand the difference without trying the program. Don't expect that you'll understand the difference in 10-15 minutes, which a lot of people think would be enough for them to understand any new idea. If you have foresaw the consequences of using the moveable / resizable elements immediately on reading the first article, then consider yourself a genius; up till now I met only ONE person, who understood this idea immediately, but that person has an exceptional mind. Usually the people begin to understand these ideas after the third try, if they are ready to read an article and try the program several times. Once too often the readers declare to themselves: "I didn't understand it immediately, then it is a pure nonsense; there is nothing interesting." I don't think that I can be listed among the most stupid of programmers (I successfully work on development of really big and complicated programs for many years), but it took me THREE months after the invention of this algorithm to understand its real value for design of different programs. These programs (from different areas, but all of them based on moveable / resizable elements) turned out to be from another world; it took me some time to formulate the rules of this world.

The samples in the demo application, which accompanies the current article, are designed to demonstrate the use of some important features in some real applications. Don't be confused by the word *demo*: applications for both articles are available not only in the form of an executable program, but with all their codes.

If you are familiar with the program that accompanied the first article ***"On the theory of moveable objects"***, then you are aware that there was not a single unmoveable object in that application. The same thing happens with the new application, which comes with this article. Only in this case there is no constantly shown information on the screen to remind you about this fact and to tell, what and how can be done in one or another form. In several forms, a short helpful information can be shown on the screen by clicking the special button, but it can be also hidden from view at any moment. The majority of the samples from the new application are the working forms from the real applications (or very close to them); you don't constantly keep on the screen the help information in the real applications.

Read again this reminder: there is not a single unmoveable object in this application.

Doesn't it look strange to you? Anyone can declare that he had spent some time on the Moon, and there is no way to refute such declaration, but with the underlined statement it is different. It is a working application, and you can check it simply by running it. And on doing it, you might feel some kind of a shock.* It will be not a surprising thing; a lot of people had the same feeling of a shock when they tried the Windows system for the first time (it was years ago; some of the readers don't even know, what was before that). By the way, the only visual difference of the Windows system from the currently used DOS was the existence of several moveable / resizable windows (in comparison with a single one and unmoveable) and the possibility of moving icons across and along the screen. That was all! And even that was a shock. Now you are going to look at the very complicated forms, in which everything can be moved, resized, rotated (if needed). I can assure you that there is much more in these forms than you can find on the first look or even the second. Keep looking and, what is better, keep trying!

The samples for this article were especially picked up in such a way as to show the real user-driven applications, working in absolutely different areas. I also tried to use the samples only from the areas, with which everyone is well familiar:

- A small application for picking out several items from the long list.
- Working with personal information (address, phones, professional status, etc.…)
- Calculator
- Plotting of Y(x) functions and a bit more.
- Some other types of plotting, including bar charts and pie charts.
- A bit of painting; just for fun.

In all these samples (with one exception, which was done on purpose) a lot of parameters can be tuned; all of them are saved for later use. Tuning of the parameters is often done via the additional forms; all these forms are also designed according with the rules of the user-driven applications. I'll write about it further on in the article.

# Data, data, data

We live in the world of data. There is much more data around us than anyone can understand, even if the powerful computers assist us in this task. We collect data, we try to visualize and analyse it. I am interested in visualizing data in any possible way that users demand, so let's look at several samples, working with data in different ways.

## *Years' selection*

Let's begin with a simple and well-known case, when you have a limited set of items, of which you need to select some subset. Such selection can be implemented for different types of objects, but for my sample I decided to stop on the years'

* Something of the same type was described by Roger Zelazny in the *Land of Chaos*; it might look like chaos to somebody from another, solid world, but there are the rules in this land, and the inhabitants of that land know the rules and how to use them at their best. After some time you begin to understand that all the rules, with which you were familiar for years, continue to work in this new land, but there are a lot of other things, which give you a lot of new possibilities. You begin to use these new things more and more; in a short period of time you can't understand, how could you live without them before.

selection. The task is simple; I don't think that anyone would have any problems on its implementation. **Figure 1** shows the view of the **Form_YearsSelection.cs** (menu position *Data – Years selection*).

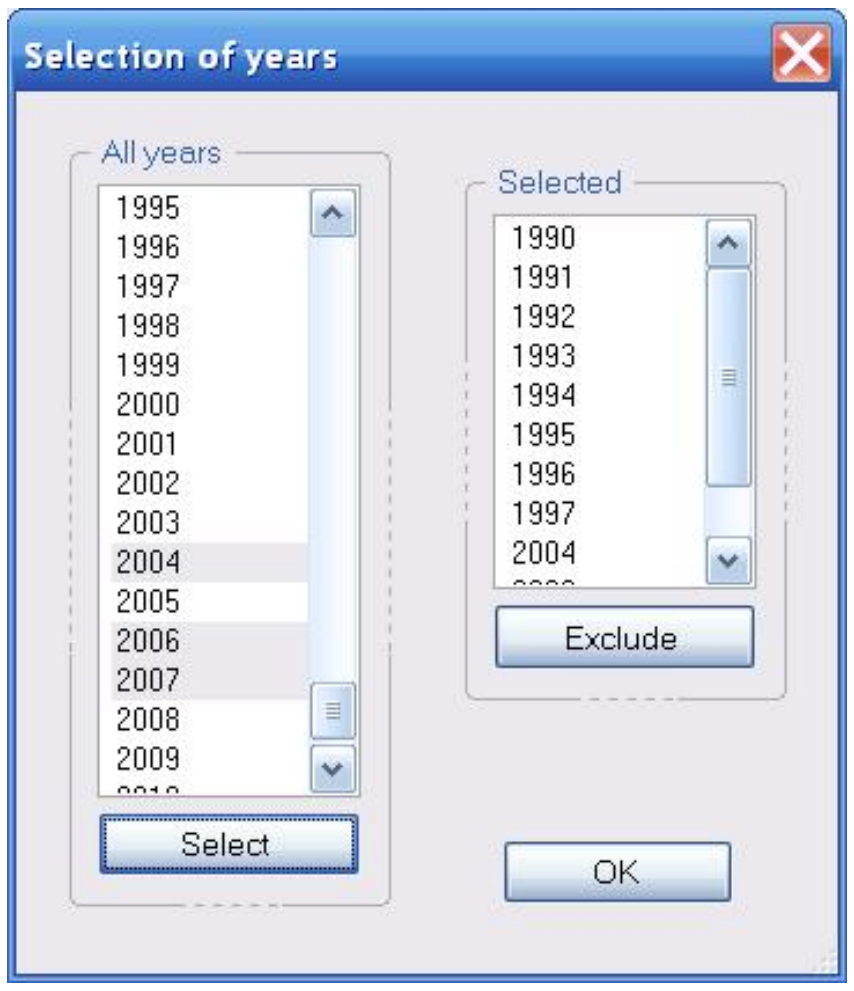

**Fig.1 Form_YearsSelection.cs**

The developers would have no problems with writing the code for this form, but the questions can be (and they certainly will be) brought up by the users on proposed design: half of the users would like to see the full list of items on the left side and the selected items on the right; another half of users would prefer those lists to be positioned in the mirror way. There is also a small percent of those, who would like to see both lists in one column; they have the right to demand such positioning. In addition, there is an old problem of the best position for OK button. Three big groups of designers prefer to place it in the right-top corner, or the right-bottom corner, or in the middle at the bottom of a form. Members of each group would never agree with anything else, except their view.

The task is so primitive, that every developer will simply design this form in the way, he personally prefers, and will ignore any other suggestions. ("The clients would have to like whatever we'll give them") As a huge respect to the users, the dynamic layout can be applied to this form; the sizes of the lists would be changed according with the used font or resizing of the form. I am sure that you can remember either designing similar thing or using it in one or another application, and there was nothing else except what I mentioned. [*]

But are you sure that this is really the right way to design even such simple programs?

Let's consider the task itself. It's only about the selection of years and nothing else. This selection works correctly regardless of the positions of all the controls and their sizes. As a designer, you are responsible for correct work of all the operations (selection, deselection, and saving); then why not to give the decision of all the controls' positions and sizes to the users and make everyone happy with the view of this form?

What has to be changed to give users the full control of the form's view?

<u>The screen elements must become moveable and resizable by the users.</u> Obvious and simple.

Why wasn't it done yet? Sluggishness of minds; I don't see any other reason. No ruling boss from the big corporation has declared the possibility and novelty of such thing, so nobody tried to do such things. It's a pity to see that the programmers' community lost the sense of imagination and the ability to think independently and only waits for the commands from the big corporation to turn left or right and to march in new direction. Until the next big turn will be announced.

So how all this moving / resizing are organized? (For those, who have read the first article and looked at the first demo application, there will be some familiar things, but I can't jump over them.)

There are three moveable objects in this form: two groups and the OK button. The groups belong to the `Group` class; objects of this class can be resized by any frame's point and moved around by any inner point. On resizing a group, the sizes of its inner elements change. This is the implementation of a classical dynamic layout, but I need to remark that this is the only form of this application, where the `Group` class is used. Further on I'll organize the groups (and they will be much more complicated groups) with the `ElasticGroup` class.

When any set of controls is united into a group, there can be one of three different approaches to organize it.

1. A group can consist of the different nonresizable parts, which can be moved synchronously around the screen. The class with such behaviour – `LinkedRectangles` – is used occasionally in this application (in other forms), but I rarely turn to this class, because I prefer to give users the possibility of resizing all the screen objects.
2. A group can be moved by any inner point and resized by any point of its frame. The resizing of the inner elements is organized as a result of the frame's resizing (dynamic layout in use). The `Group` class belongs to this type; there are more samples of the similar classes in the **Test_MoveGraphLibrary** application. Though all such classes perfectly fit with the ideas of the user-driven applications, I prefer the groups of the third type.
3. In this type of group, all the inner elements can be moved and resized individually; these elements do not influence or put any restrictions on each others movements. The frame around the elements can be shown or not, but in any case it doesn't rule the positions and sizes of the inner elements; on the contrary, the group's area is defined by the united area of all the elements. The synchronous movement of all the elements can be started by any inner point or by the frame. The `ElasticGroup` class is the best representative of this type of groups.

[*] If you think that I am not correct, let me know about at least one application of such type, in which users get a chance to decide about the relative positions of these lists. This is such a common task of data selection that you would see something similar often enough. I would like to know, if there was at a list a single application, which differs from what I have written.

There will be more on the third type of groups in the further samples, but now let's return to the case of the **Form_YearsSelection.cs**, which uses the `Group` class.

There is a mover object (of the `Mover` class), which supervises the whole moving / resizing process. To be involved in this process, the form's objects must be registered with this mover.

```
mover .Insert (0, groupAll);
mover .Insert (0, groupSelected);
mover .Insert (0, btnOK);
```

The three standard mouse events are used for the whole process of moving / resizing; the code of these events can't be simpler than in this form.

```
private void OnMouseDown (object sender, MouseEventArgs e)
{
    mover .Catch (e .Location, e .Button);
}
private void OnMouseUp (object sender, MouseEventArgs e)
{
    mover .Release ();
}
private void OnMouseMove (object sender, MouseEventArgs e)
{
    if (mover .Move (e .Location))
    {
        Invalidate ();
    }
}
```

You can see a single call to the mover's methods in each of three mouse events, and that is all that is needed. In further samples (they are much more complicated) you'll see some additional lines in the code of these events; not many lines, but some. All those extra lines are added for some auxiliary things. If in any of those samples you'll take out (comment) all the additional lines, you'll lose some features of those forms, but all the objects (regardless of their number and types) will be still moveable and resizable. These three calls to the mover (one line per each event) are all that is needed to move and resize the objects.

The task is a simple one; the implementation is also simple, but even this sample allows us to make the first acquaintance with the basic rules of the user-driven applications.

**<u>Rule 1</u>. <u>All the elements are moveable.</u>**

There are no exceptions; all the objects, regardless of their shape, size or complexity must be moveable. If for some object you don't see a good solution in an instant, THINK a bit more, and you will find a good solution. Users have to get the full control of an application (form), which means that each and all objects must be under their control. Users are going to use an application at the best level to fulfill their work; the moveability of the elements increases their chances to do this work exactly in such a way, as they want, so give them this chance. If you'll decide that you are giving them nearly everything, **but** this and that, then they will be bumping into these hillocks on the road all the time. With an adequate thought about you, as a designer.

**<u>Rule 2</u>. <u>All the viewing parameters must be easily controlled by the users.</u>**

Rules 1 and 2 are the projections of the same full users' control over a program on the different sets of parameters. Rule 1 deals with the locations and sizes; rule 2 - with the colors, fonts, and some auxiliary things.

The rules that are mentioned here are strongly related with each other and always used together. But because the accompanying program is a Demo application, I purposely violated the rule 2 in this particular **Form_YearsSelection.cs**, and there is no tuning of parameters in this form. It is implemented and demonstrated in details in all other forms, which are much more complicated, but not here. After reading several more pages and being familiar with other forms of this Demo program, return back to this simple form and estimate your feelings in the form without such tuning of all the viewing parameters. I think that this will show you more than anything else the importance of this rule. (There is also no tuning for the groups in the first Demo application; that article was only about the algorithm, and I didn't want to jump far ahead.)

**<u>Rule 3</u>. <u>The users' commands on moving / resizing of objects or on changing the viewing parameters must be implemented exactly as they are; no additions or expanded interpretation by developers are allowed.</u>**

Changing of the viewing parameters by the users is not an unknown thing and is implemented in a lot of applications. But in the standard applications, especially those that are built on the ideas of dynamic layout, users would change one parameter, for example, font, and a lot of related things would be changed automatically, because this is the nature of dynamic layout. With the user-driven applications you have to stop thinking in the way like this: "You changed the font. I am smart, I know what you really wanted to do, so I'll do it for you: I'll adjust the sizes of this, that, and that object. Be happy, because I save you several clicks." This is an absolutely wrong way of design for user-driven applications. The developer <u>must not</u> interfere with the users' commands and add anything of his own.

It can be a bit strange at the beginning of new design to control yourself and not to add anything of your own to the users' commands. You may be a designer with many years of practice, you really know, what must be done to make the view of an application better in this or that situation. But this is another world; if you gave users the full control of an application, it must be really full, so you don't leave anything for yourself as a second control circuit. Whatever is gone is gone.

And eventually you'll find that nobody needs your even excellent experience on adjusting the forms to their needs. Where you have to apply all your skills in design (the higher – the better) is in construction of the default view of every form. The highest credit to your skills is the big percent of users, who would not change anything at all, but work exactly with your proposed design. And yet, the possibility of all those moving, resizing, and tuning must be there for users to try at the first wish.

**<u>Rule 4</u>. <u>All the parameters must be saved and restored.</u>**

Saving the parameters for restoring them later and using the next time is definitely not a new thing and is practiced for many years. But only the passing of the full control to the users with the significant increase of the number of parameters that can be changed, turned this saving / restoring of parameters from the feature of the friendly interface into a mandatory thing. If a user spent some time on rearranging the view to whatever he prefers, then loosing of these settings is inadmissible. And as the full users' control means the possibility of changing any parameter, then the saving / restoring of all the parameters must be implemented.

It doesn't mean that you'll have to write many lines of code, if you are going to organize a form with dozens of objects inside. Beginning from the next sample, we'll deal exactly with such complicated forms, but you'll see that the process of saving / restoring is not boring at all or time-consuming.

These four rules were born in different ways. Rule 1 – the main rule of user-driven applications – fixes in words the whole process of transformation from the standard applications into something totally different, based on moveable elements. It took me several months to understand the entire process; the understanding came step by step, and on each stage I had no idea, what would be the next. But it turned out to be very logical; I'll talk about it further on, while writing about functions and scientific / engineering applications, because this is the area, where all these things started. Rules 2, 3, and 4 became obvious to me and were formulated later, when I started to design more and more user-driven applications. With all these main rules formulated, let's turn to the next sample and see, how these rules work in a real case of a form with a lot of controls.

## *Personal data*

A lot of forms in different applications are constructed of controls. A set of controls in each particular case depends on the goal of application and the designer's decision; such set can include one, several or a lot of controls. There are exist between 20 and 30 often used types of controls plus variations on the theme. Controls can vary in sizes from small to big; they have a lot of parameters, which allow to show them in many different views, but for using them in design of user-driven applications we can formulate such interesting rule:

**Any form (dialog) without graphical objects can be constructed of elements of only three different types:**

- Individual controls
- Controls with comments
- Groups of controls

The formulated rule is a recursive one, as any group of controls can be constructed of the same three types of elements. The `ElasticGroup` class is an excellent demonstration of turning this rule into the code. This class is widely used in the majority of the further samples; it is also used in design of many auxiliary forms, which I am not going to mention here in the article, but which you'll immediately see, when you start looking on the working application.

Controls significantly vary in their appearance and behaviour. From the users' point of view, there are no similarities between a `ComboBox` and a `Panel`, or between any type of `Button` and `ListView`. But from the point of moving / resizing the controls, their behaviour is identical, and the `FramedControl` class, which was described in the first article, works perfectly with any control. I have already explained in the former article that whenever we talk about an individual control and its moving / resizing, in reality there is an invisible `FramedControl` object, which wraps this control and provides all the needed procedures.

A pair "control + comment" is widely used in the applications' design; in the world of unmoveable controls, the "comment" part of such pair is often represented by another control – a `Label`. In the world of moveable / resizable elements, it's much better to have the "comment" part of the pair as a graphical object (painted text). Such a pair "control + graphical textual comment" is represented in my design by the `CommentedControl` class. The moving / resizing of such pair can be described by several rules.

- The moving / resizing of the control of such pair are identical to the same operations with the individual controls.
- On any moving / resizing of a control, the comment preserves its relative position to the control.
- The comment can be moved (and rotated) individually. There are no restrictions on positioning of the comment in relation to its partner – control except one: the comment can not be totally covered by the control, because in such case it will slip out of the user's control. To avoid this situation, whenever the comment is moved around and released while totally covered by its associated control, then it is forcedly moved slightly outside so that it becomes visible and accessible. I want to underline that this enforced relocation is used only when the comment is closed from view by its own associated control, and not by any other.

The groups, which are used in the further samples, belong to the `ElasticGroup` class; objects of this class can consist of any number of the individual controls (`FramedControl` objects), controls with comments (`CommentedControl` objects) or other groups of the same `ElasticGroup` class. The name of the class underlines its overall behaviour: such group always surrounds all its inner elements regardless of their sizes and positions. Let's see, how the `ElasticGroup` class can be used for design of really complicated forms.

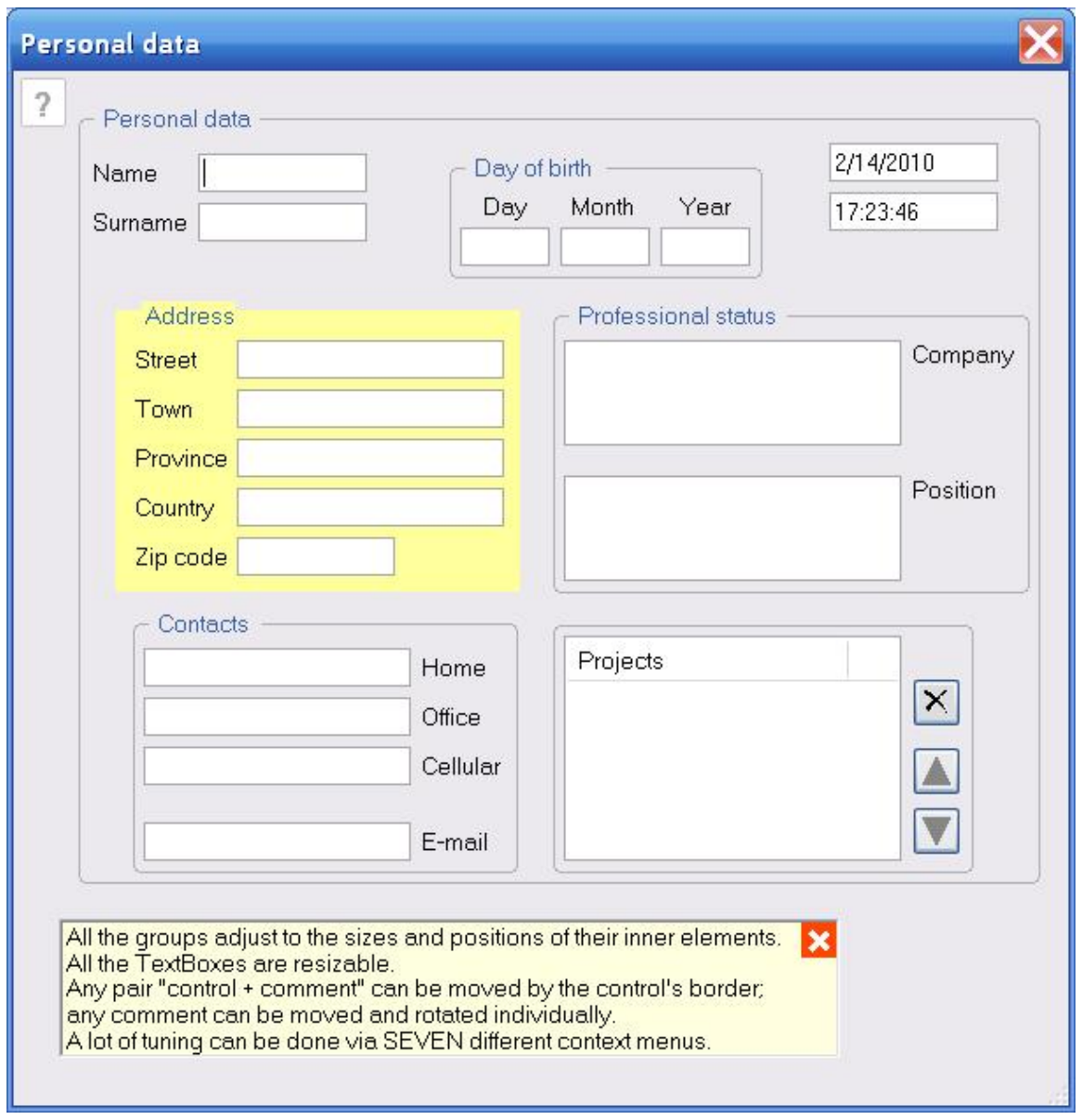

**Fig.2** `ElasticGroup` objects can include other groups of the same class, thus allowing to construct very complicated forms

**Figure 2** shows a view of the **Form_PersonalData.cs** (menu position *Data – Personal data*), which consists mostly of different groups. With the default set of parameters, any `ElasticGroup` object is indistinguishable from the standard `GroupBox`, but the moment you start doing anything in this form, you'll immediately understand that this is not an ordinary one.

This is not an artificial sample to demonstrate one or another discussed feature. This form deals with the collection of data and the situation, which is familiar to everyone throughout his life, so everyone can make his own decision about the usefulness of the proposed design on the basis of the moveable / resizable elements. Certainly, each of us deals with the collection of personal data only occasionally, but there are people (from HR departments), who work with similar forms day after day, and it would be very interesting to hear their opinion.

The **Form_PersonalData.cs**, which is obvious from its name, deals with the personal data. Depending on the case, it can include only a small piece of data, or might require to include more, than is shown here, but you are free to add any other needed groups of information (all the codes are available).

The `ElasticGroup` class has a significant number of constructors; there are also nearly 20 different constructors for the `ElasticGroupElement` class. Together they provide such a variety, that you can find the easiest one to use for any type of the needed group either it consists solely of the elements of one type or of their combination. In the shown form, the biggest (outer) group includes two `FramedControl` objects to show date and time, two `CommentedControl` objects to show name and surname, and five inner groups. Overall there are 23 controls in this form. I wouldn't call it an extremely complex form (you will see more controls in the next sample), but at the same time I think that the majority of readers rarely had a chance to design a form with such a number of controls. Of all the elements only the *Name* and *Surname* can't be hidden (I don't see any sense in hiding this part of personal data); everything else can be used in any combination.

Any `ElasticGroup` object consists of a number of its inner elements; there can be many variants of initializing these inner elements; I tried to demonstrate different possibilities.

For an element, consisting of a single `CommentedControl` object, I construct this object beforehand.

```
    CommentedControl ccName = new CommentedControl (this, textName, Resizing .WE,
                                                  Side .W, "Name");
```

For an inner group, consisting of several commented controls, an array of such objects is prepared.

```
CommentedControl [] ccsDOB = new CommentedControl [] {
        new CommentedControl (this, textDay,   Resizing .WE, Side .N, "Day"),
        new CommentedControl (this, textMonth, Resizing .WE, Side .N, "Month"),
        new CommentedControl (this, textYear,  Resizing .WE, Side .N, "Year") };
```

With the sets of inner elements already prepared, the construction of the big group is easy.

```
groupData = new ElasticGroup (this, new ElasticGroupElement [] {
                  new ElasticGroupElement (textDate, Resizing .WE),
                  new ElasticGroupElement (textTime, Resizing .WE),
                  new ElasticGroupElement (ccName),
                  new ElasticGroupElement (ccSurname),
                  new ElasticGroupElement (this, ccsDOB, "Day of birth"),
                  new ElasticGroupElement (this, ccsPhones, "Contacts"),
                  new ElasticGroupElement (this, ccsAddress, "Address"),
           new ElasticGroupElement (this, ccsProfessional, "Professional status"),
           new ElasticGroupElement (this, new Control [] {btnDelete, btnMoveUp,
                                                  btnMoveDown, listProjects }) },
                              "Personal data");
```

Now it's time to register our objects with the mover (only then they will become really moveable / resizable), but several words before this. In the previous sample it was enough to use a simple `mover.Insert()` method, but that would work correctly only for the primitive objects or objects, considered by mover as a single element. I have already explained in the first article that usually two independent, but working in collaboration methods are used for correct registering of complicated objects and organizing the correct mover's queue in a form with a lot of elements.

- Each class of complex objects has its own `IntoMover()` method, which guarantees that any object of this class, regardless of the number of its constituents, is registered correctly.
- The `RenewMover()` method belongs to the form and guarantees that all the form's objects are registered fully and correctly, regardless of their number and possible change of order during the work. This method never deals with the individual parts of the complicated objects, but only with the `IntoMover()` methods of those objects.

Now we can easily write the `RenewMover()` method for the **Form_PersonalData.cs**.

```
    private void RenewMover ()
    {
        mover .Clear ();
        groupData .IntoMover (mover, 0);
        if (info .Visible)
        {
            mover .Add (info);
        }
        mover .Insert (0, btnHelp);
    }
```

As you can see from this code, the registering of the whole outer group is done with a single code line. There is a lot of possibilities to change the set of visible elements of the group or their visual parameters; these changes would often require to call the `RenewMover()` method, which will always guarantee the correct order of elements in the mover's queue.

The `OnMouseUp()` method of this form contains some auxiliary lines (in comparison with the previous sample), but they are only to select one of the context menus. There are seven different context menus in this form, which can be called on different types of elements or at any empty place.

Let's check in details, how the four previously mentioned rules of the user-driven applications work in this particular form. Those four rules are not absolutely independent, but, on the contrary, they are strongly related with each other, so some of the further explanations, linked with one of the rules, can be easily moved into the part, "belonging" to another rule. There are a lot of different controls, combinations of controls, and classes in this form, so we have a chance to analyse different situations.

**Rule 1. All the elements are moveable.**

Yes, ALL the elements are moveable. I wouldn't even try to predict the number of different users' opinions about the best view of such a form. I think that it would differ not too much from the number of users. Or it would be significantly higher than the number of users, because my experience shows that when there is a chance to rearrange the view of the form (application) at any moment, then each user has different opinions about the preferable view each day of the week and also depending on the time of the day and the weather outside. The moveability of each and all parts of the form allows to rearrange it in an arbitrary way.

The rules of "good design", which can be found in many books and manuals, declare that if you have in a form several controls with the comments, then all the comments must be placed on the same side of the controls. You can see from **figure 2** that I have not slightly violated, but absolutely ignored this rule. I think that with the spread of user-driven applications a lot of currently used rules of design, even those that are looked at as axioms, will be ignored, forgotten, or revised. Another world, different rules. The majority of currently used rules were declared 20 and plus years ago, when everyone was working with a single screen. Now we have much bigger screens; a lot of people use several screens. But the biggest blow to those rules comes from the moveability of the screen elements.

I don't see anything annoying or strange if in one group the comments are positioned on one side of the `TextBox` controls, while in another group they are positioned differently. Look at the central part of **figure 2**; there are two groups, which look like opposite pages of the opened book. Why not to have the comments to these groups on the outer sides of them? It's good for the books, why is it going to be wrong for the screen? By the way, in the realm of old design (fixed elements) you have no chances to check, if the announced rule is right or wrong. As a designer, you would construct the forms according with this rule; maximum that you would allow yourself is to try either the left side positions or the right side positions for all the comments, but nothing more. (It is my assumption; I can be wrong.) As a user, you would have to work with what you had been given, and in nearly all the programs this rule is implemented. Only now, with all the elements easily moveable around the screen, you get the chance to check that rule for yourself and to decide about its usefulness or not.

There is another thing, related to the same problem of multiple `TextBoxes` with comments: the alignment of all those comments. The preference on this item is so individual that I don't remember even reading about any rule; the comments' alignment was enforced by the designers without any second thought. ("You would have to like what I like.") I wasn't arguing about the comments alignment years ago, I wouldn't do it know, when I give you an instrument to change this alignment easily, quickly, and in any possible way. If you have a very specific view on the good looking design, you can place those comments even in the chess order around the `TextBoxes`. I'll not say a word against it. You took the application; you work with it; you change it in the way, which is the best for you.

Some people can say that these are the minor items, which can be simply ignored. I don't agree with this. And I think that the users of such programs, especially those, who have to work with similar applications many hours a day (HR people), would not agree that these things can be simply ignored.

Several more remarks about this form.

The highlighted group at **figure 2** represents the standard set of controls to deal with the address information. While the order of controls in this group is natural for western countries, it is absolutely unnatural for other parts of the world, where the address is often written in the opposite order, beginning from the country information. With all the elements being moveable, it will take a couple of seconds to rearrange the view to whatever the user would like it to be (to what he got used for many years). I can't think out any other good solution for this problem. Certainly, you can hard code several variants plus the selection of the one, which is needed. It is definitely a lot of extra work, and are you sure that you would put into code all the needed variants? Or you can simply ignore the specific local preferences in the address information. (Let them work with what is given.) You can decide for yourself, if it is a good solution or not. What would you prefer, if you have to work with such data day after day: movable elements or a fixed design?

I have already mentioned in the first article that one specific type of moving the elements to rearrange the whole view is widely used in the user-driven applications: you can simply move currently not needed elements out of view across the borders of the form. There is an easy way to put the needed level of restrictions on such movement, so you'll never lose your objects, if you don't want to do it. The temporary relocation of the objects across the border of the form is not a trick in the user-driven applications, but a very reliable, often used, and very easy and quick way of adjusting the view of applications. This method can be used with any stand alone object (it can be a single control or a whole group with a lot of elements inside), but it's not the good idea to try this technique with any inner part of the surrounding group. If it is a part of the `ElasticGroup`, then this group's frame will go after its runaway element not looking at the distance.

For the `ElasticGroup` class, there exists much better solution: any inner element can be turned into invisible and returned back to life only when it will be needed again. This technique can be used at any level. For example, consider the group *Address*. There are a lot of countries, in which no *Province* information is required (no intermediate level of addressing between the *Country* and the *Town* is used). You don't need to keep at the screen the unneeded controls: call the menu on this "control + comment" pair and hide it. If you work only on the data of local people, then you can get rid of the *Country* pair also. If you don't need a whole inner group, for example, the *Projects* group, it can be made invisible in the same easy way through its own context menu or the menu of the big group. Any invisible element can be returned back to life via the context menu of its parent (surrounding group).

The moveability of the elements adds so many new things to the design of applications! I work on the design of really big and sophisticated programs for more than 30 years; new languages and faster computers opened new possibilities from time to time, but it was never a revolutionary step, but only some improvement to the already known procedures. With the invention of the algorithm for turning elements into moveable / resizable, the situation is different. There are no recommendations on what can be done in one case or another; you go step by step in exploring the new continent, and often you find the things, which you couldn't even imagine a month or two ago.

The `GroupBox` class is known for "centuries"; the group's title can be shown either on the left, or on the right. I have designed several classes to move the groups of elements; some of those classes are demonstrated in the **Text_MoveGraphLibrary** application. If any of these groups has a title, then its positioning is described by the `StringAlignment` enumeration, so there are three possible variants: left, center, or right. The `ElasticGroup` class was originally designed with the same type of title's alignment, but one day I looked at it from the different angle, and now it has a title, which can be moved along the upper border of a group and placed in any position. Such change didn't require any significant additional work, but only different look. This often happens with the design of moveable objects: think about them not as a slightly different copy of the well known object, but as a representative of another world, in which everything is possible. I'll slightly jump ahead, because this object is shown at the next figure, but the same thing happened with the `TrackBar` control.

Design on the basis of moveable elements may put in front of you some questions, which would be impossible to imagine even a month before. Throughout the last three years I was designing more and more complicated programs, consisting of moveable elements; the number of moveable elements on the screen was growing all the time according with the users' demand. Then users ran into situation, when, with the big number of moveable screen objects, the wrong object was accidentally moved instead of the right one. Users required an easy way of turning any object (simple or complicated) from moveable into non-moveable and back again.

In case of the **Form_PersonalData.cs**, this problem of accidental movement of the wrong object becomes obvious with some of the groups, in which the group's area is packed with its inner elements. For example, look at the *Address* group (**figure 2**); there are five pairs "control + comment" in it. Suppose that you positioned all the controls and their comments in the way you want them to be; all these elements are moveable, so you have no problems with rearranging the group. Now you want to move the whole group somewhere else; to do this, you must be a bit careful not to press the mouse anywhere near the inner controls or comments, otherwise instead of the whole group some inner element will start moving, and you don't want to do it. On opening the context menu for this group, you see one line, which allows to change the moveability of the group's inner elements; depending on the situation, this line is either *Fix group's elements* or *Unfix group's elements*. If you fix the elements, then you can move the group easily and not being afraid of accidentally moving some inner element. The inner elements are still recognized by the mover, and the appropriate context menu can be called on any of them, but individually they are not moveable, until you'll unfix them.

From the programming point of view, this switch between moveable / unmoveable elements is easy to implement. When you design a class of moveable graphical objects, you write its `DefineCover()` method, in which you define the cover as a set of nodes (`CoverNode` class). Each node, among other parameters, has a parameter of the `MovementFreedom` enumeration. If this parameter is set to `MovementFreedom.Freeze`, then this node can't be used for moving, but it is still recognized by mover, so, for example, a context menu can be called on this element to change its view. (More about `MovementFreedom` was written in the first article.)

The **Form_PersonalData.cs** is based on the `ElasticGroup` objects; this class has two different properties to deal with the moveability. One of them – `Movable` - sets the moveability of the group and all its inner elements. This property includes the call to the same property of the base class (`GraphicalObject`), which includes the call to the `DefineCover()` method, so the cover always corresponds with the current (needed) state of moveability.

```
new public bool Movable
{
    set
    {
```

```
                ElementsMovable = value;
                base .Movable = value;
            }
        }
```

The `ElementsMovable` property sets the moveability of all the inner elements, but not of the group itself; thus you can achieve different combinations of moveability for the group and its inner elements. I suppose that usually the *Address* group would be organized in the needed way; then its inner elements will be turned into unmoveable, but the group itself will be still moveable. But if you wish, you can fix the elements of the big group, thus making the *Address* group unmoveable, and then declare the inner elements of this group moveable. I don't see the situation, when I would need such a combination, but… I want to emphasize again. I am only a designer and I give users an instrument to work with. As a designer, I have to provide them with a good working instrument, and that is all. Not a bit more; I have no rights to tell users, how they have to work. They will decide how to use this instrument. I have no rights to interfere with their work and put any restrictions on it. It is a *user-driven application*, not a designer-driven.

As a user, do you prefer such design, or do you think that it is too much of a burden for you to have all these possibilities? Do you want to rearrange an application in the way YOU want it to be, or do you want to work inside my vision of situation, slightly decorated by dynamic layout?

One more remark on moveability of each and all. Complex applications have a lot of screen objects. These objects are not only moveable, but a lot of them are complex objects, which include individually moveable parts. To all these individual and synchronous movements, users can add the moving of the arbitrary groups of objects (the sample is described further on in *Calculator*). After series of movements, user can understand that the default view was much better then he organized himself (you are a very good designer!). It's a good idea to give users a chance to reestablish with one click a default view of a group or of a whole form. In the standard applications, you don't even think about such problem, because those applications always have a default view. For the user-driven applications, it's not another rule, but simply a sign of good design. A sign of your respect to the users of your programs.

**Rule 2. All the viewing parameters must be easily controlled by the users.**

There are several parameters, which influence the view of the forms (colors, fonts, sizes of elements, and distances between them), but because these parameters can be defined individually for each element, then the number of possible variants is huge. There are two common ways to set the parameters of any object: use the context menu or call an auxiliary tuning form. The `ElasticGroup` objects are used throughout this application; each object of this class has around 15 tuneable parameters. Only on rare occasions I organize the tuning of a group via a set of positions in its context menu; more often you open the context menu and click the line *Modify group*; then the special **Form_ElasticGroupParams.cs** is opened for tuning (**figure 3**).

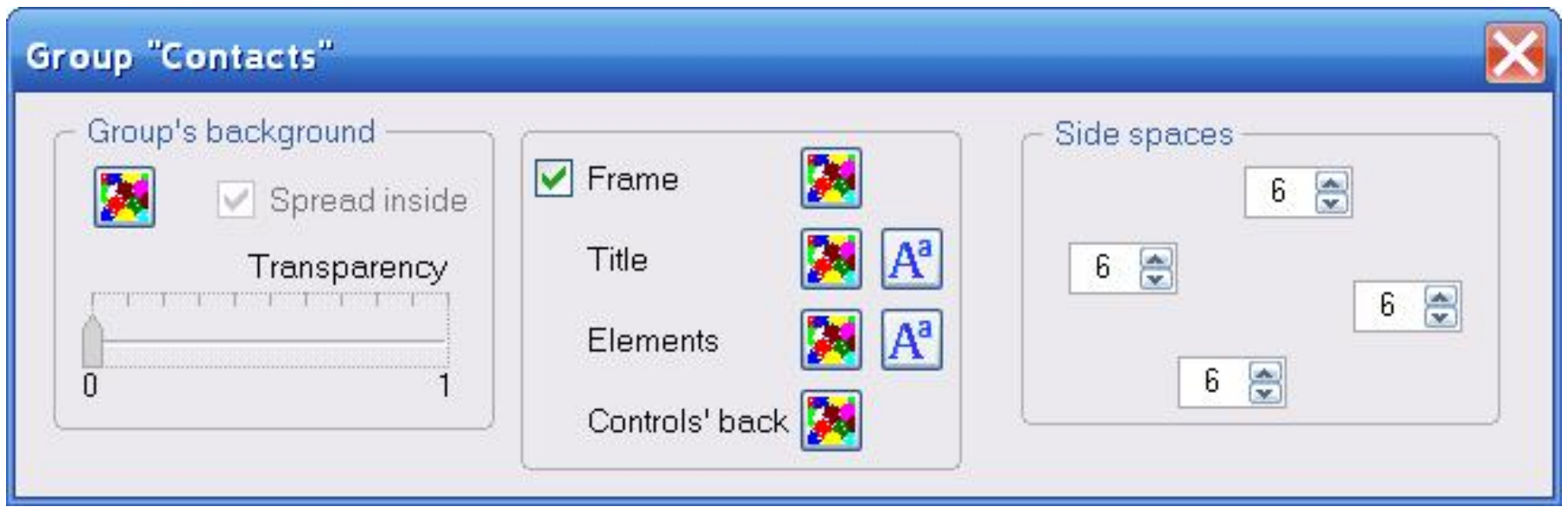

**Fig.3 Form_ElasticGroupParams.cs**

When the transparent panels were first introduced to the public, a lot of users decided that that was the goal of the programmers, to which they were going for many years, and that there would be nothing even comparable with such an achievement from now and forever. I never thought that such feature even worth talking about, but anyway, if users want, they can change the level of transparency for any group.

Group is usually surrounded by the frame, which not only makes obvious the group's bounds, but, what is more important, marks the set of controls that constitute the group. But the frame is not the only way to fulfill both goals; coloring the group's area can do it not worse. Drawing the frame of a group and coloring its area are independent, so they can be used even simultaneously. Users have all the choices.

When the alignment of the group's title is set via the context menu, then I usually give three choices: left side position, center, or right side. But there is another solution, which is very easy to implement, but which is not shown in the tuning form, because this parameter now became the part of … moveability. The title of the `ElasticGroup` object is moveable and can be placed anywhere along the upper border of the group. It's definitely the designer's choice to organize such a group, which he would like to see in his programs; a title in such a group can be unmoveable, or the title might have several predefined positions, or its arbitrary positioning can be allowed. I don't want to set any rules or limitations on the moveability of objects or their parts. I want to show that when you start designing in this world of moveable / resizable elements, you find that a lot of new things are possible and very useful; the things, which were unthinkable before.

When you are tuning a complex group, which includes other groups as its inner elements (like *Personal data* group at **figure 2**), you can decide for yourself about spreading the background color of the outer group on the inner groups or not.

With so many tuneable elements in the form, it is not a bad thing to have an opportunity of using one of the elements as a sample for others. I didn't include such a thing into this particular form, but it is implemented in further samples (for example, look for the spreading of comments' parameters in the **Form_PlotsVariety.cs**).

The **Form_PersonalData.cs** has a lot of elements (**figure 2**); each element has a number of tuneable parameters, so there are a lot of possibilities for users. This variety leads to the same questions and concerns, which I hear from time to time, but only from the people, who do not use such type of programs and at best only look at them for several seconds to give out their extremely professional and authorized opinion. More often than not these people do not even look at the program (they think that they are too busy even for this, and anyway, they know everything without looking at the programs); it's enough for them to read an article along the diagonal. Here are the standard remarks from such people; if you jumped to the same ideas, then you are definitely not alone.

**Remark 1** Users don't need all this setting of parameters; they need only the working program and will be happy with it.

My answer The work of the program does not conflict with the possibility of setting the parameters. All the tunings do not change the goal of the program, but only make it much more flexible and provide a lot of possibilities to rearrange the view of a program to whatever each user would like to see.

**Remark 2** Too many parameters for tuning; users will be lost in them.

My answer I agree that the number of tuneable parameters can be really big, but this is the reality of the big programs. By not allowing users to change those parameters, you do not decrease their number; all the parameters are still there, but in the standard applications they are set at the developers wish, and that's all. Give users an instrument for easy changing of ALL these parameters and let users decide, when, how, and to what extent they would like to use this instrument. The decision must be given to users and not stolen from them.

Users do not change the parameters all the time. Usually they set them at the beginning of the work and then there is only an occasional additional tuning, when something needs to be changed. But at the beginning it's often the only way to get the good looking application. I spend a lot of time, trying to design the complicated forms in my applications in really good view. Then I give these programs to colleagues, and look with horror at the initial view of the same programs on other computers. Even if I would try to, I wouldn't have a chance to destroy the views of my forms more than I see, when they are used for the first time on different computers.[*] From my point of view, the modern PCs are like a bus, in which each passenger (each service program) has his own steering wheel. Nobody knows whose wheel is the main one at one moment or another; it looks like they take this role at random, but definitely there are several of them.

With the applications, constructed of moveable / resizable elements, I am not worried any more about this strange effect on different computers. When a user-driven application starts at any computer for the first time, it may has not the best view, but each user gets a chance to redesign any form to a good looking one in seconds. This may happen (I am sure it will happen occasionally) with the forms from the accompanying application, but the figures from this article can show you, what I prepared and what I would like you to see. You can rearrange each form to the identical view, or you can change them in any way you want. It's a user-driven application, in which the ideas are the main things, and the views are under your full control.

The "experts in design" deny the idea of giving the full control to users as the biggest heresy; they don't want neither to discuss the possibility nor even hear about it. I never heard a single complaint about this thing from the users, but an outcry started from those, who teach for years, how to design the good looking system. It looks like those people simply try to defend their monopoly on the sacred knowledge; nothing else. What are they afraid of? Of the fact that users can organize the applications' view in the way they really need? But what is the problem with this? Programs must have the best design (view) for each user personally, but not for the average user, as a designer imagine him. (I hope you understand the difference between these two things.)

The idea of moveability penetrates all the layers of design; the moveable big parts allow to change quickly the whole view of an application or a form; the same feature of the smaller or tiny parts allow to reach the aesthetic perfection in parallel with the most effective work of applications. The good sample of such use of moveability is demonstrated in the mentioned auxiliary form (**figure 3**). To set the transparency of the modified group, the `TrackBar` is used; it is definitely not the standard `TrackBar` from VisualStudio, but its graphical analogue.

[*] It's incredible even to imagine, what this Windows system can do to the good looking application. The developers of this system must have the motto, which was once pronounced by the former Russian prime-minister: "We wanted to do the best, but the result was as usual…"

I have already explained in the first article my point of view that the substitution of the standard controls with their graphical analogues would be a huge benefit for form's design. Here is an excellent sample of such substitution. The idea of a `TrackBar`, which allows to select any value from some range, can be very useful in many situations, but the implementation of this control in VisualStudio is so clumsy that after each attempt of using the standard `TrackBar` I had to reject it and turn to something else. But the idea of this control is perfect, so I designed the graphical `HorTrackBar` class, which fits perfectly with all other elements of the user-driven applications and with the rules of such applications. This track bar:

- can be moved around by any inner point;
- can be resized by the left and right borders;
- may have any number of arbitrary positioned comments, which move synchronously with the track bar, but can be also moved individually, if it is allowed. The last thing is set individually for each comment, so, for example, the end values of the shown track bar are fixed at the ends of the bar, but the title *Transparency* can be moved to any relative position you would like to.

Such moveable / resizable track bar behaves as any other element, so its moving / resizing have the same effect on the frame of the surrounding group, as of the button or the `CheckBox` in the same group.

Font of the elements is only one of the visibility parameters; according with the *rule 2*, control over the fonts is also passed to users. But fonts' changing has more consequences, than changing of other parameters; this action is strongly linked with the *rule 3*.

**Rule 3. The users' commands on moving / resizing of objects or on changing the viewing parameters must be implemented exactly as they are; no additions or expanded interpretation by developer are allowed.**

The dynamic layout is used in applications for years and throughout these years we got used to some of its remarkable features. The dynamic layout is some kind of designing philosophy, but we, as users of different programs, deal with the implementation of this philosophy in each particular case. This can differ a lot from application to application, because it depends on the interpretation of the ideas of dynamic layout by each designer. Our expectations can be different from the developer's view; this is the cause of many problems. I still don't understand why the command to change the font's size has to change the form's size also. I didn't ask for it, I want to see the same information at the same place, but in bigger letters; that is all. Why somebody decided, even without asking me, that I would like to enlarge the form and controls? From my point of view it's really obtrusive, and I can't understand, why it is continued to be called a friendly interface. (It's exactly like washing my car at my expense at the gas station, if somebody would decide that any filling of the gas tank must be accompanied by the total car wash. I really wonder how many of the readers would call such a gas station friendly.)

The user-driven applications give the FULL control to the users, so the designer of such applications has no rights to add anything of his own to the users' commands. Thus the command on changing the font of a single element or a group would do exactly what was ordered and not a bit more, but… the result can be slightly different from our expectations, based on all our previous experience.

- Let's change the font of the comment, belonging to the `CommentedControl` object. The comment of such object belongs to the `CommentToRect` class, which is derived from the `TextMR` class. These objects are positioned by the central point; on changing the font, the anchor (central) point is not moved, but the text's area changes its sizes by expanding in all directions or shrinking. If the comment is positioned close to its related control and you significantly increase the comment's font, then chances are high that the enlarged comment will be partly covered by this control; the text will need some moving to reestablish the good view of this pair.
- When you increase the font of a whole group, the result depends on the type of controls in this group. Some controls will preserve their sizes and will not organize any troubles for you; others will automatically change their sizes. If several text boxes are positioned close to each other, then chances are high that after the font's increase they will overlap; you'll have to move some of the controls to restore a good view.

While working with the user-driven applications, you can occasionally observe such or similar annoying results of the font's change, but in reality such problems will occur rarely enough. First, the significant increase of the font is a very rare thing; for the small changes of the size the effect will be not as bad as I mentioned. Second, you don't change the font too often. Usually you change it once, make some needed movements of the elements, and after it there will be no requirements on changing the font.

When you try to introduce and use something new, you have to take into consideration the features of the elements, on which you and others build the applications. I want to cut the strong link between the control's sizes and the size of the font, which this control uses. I also don't want to include into my applications any actions, which are still out of users' control. For this reason, I never change the font for the whole form throughout the form's `Font` property, which, depending on the form's `AutoScaleMode` property, might also change the form's size. Instead I change the font of each control on an individual basis; the consequences depend on the type of those controls. For example, any single lined `TextBox` will change its height according with the font's size, and you have no chances to do anything with it; this reaction is fixed deep inside the system by the controls' developers; programmers have no access to it and no control over such action. The majority of the `TextBoxes` in the **Form_PersonalData.cs** is of such a type (the `Multilne` property is set to `false`); the width of such element is under users' control, but not the height. However, you can give users the control over the height of such element, but only if you set the `Multiline` property of the `TextBox` to `true`; such technique is used in *Calculator*, which is discussed further on. Some negative effect can be observed on the significant increase of the font for such `TextBox`: at first moment the increased symbols are seen through the same area of a control only partly, until the users will resize this control with a mouse. Everything comes with a price.

About the violation of *rule 3.* I would prefer to have no exceptions from rule 3, but at the moment there is one, which is related to the `CommentedControl` class. User-driven applications can use a lot of different classes, and similar situation might occur somewhere else. Any `CommentedControl` object consists of two parts: control and comment. Any moving of a control causes the synchronous movement of its comment, so there are no problems at all. Comment can be moved individually and released at any place. Controls are always shown atop all the graphical objects; when the comment is moved across the area of any control, it is hidden from view. What will happen, if a comment is released at such a moment, when it is totally closed from view by a control? If this is not the control of the same pair (of the same `CommentedControl` object), then the comment is left, where it was released. User can move the control aside and open an access to the comment. If the control belongs to the same `CommentedControl` object, then there would be no way to relieve the comment; it becomes invisible and inaccessible forever. To avoid this situation, the system violates *rule 3* and in this particular case pushes the comment slightly from underneath the control, thus making the comment accessible.

The described situation happens at the moment, when mover releases an object (a comment or a control), so in this case the mover is responsible for checking the possibility of such situation and avoiding it. But similar situation can happen as a result of the things, not related to mover in any way, and then a designer has to think about it. Consider a case of a `CommentedControl` object with the comment, using some big font and positioned mostly under its control with only a tiny part of the comment in view. Then you call a context menu on this comment and order the significant decrease of the comment's font. As a result, the area of the text will shrink and totally disappear under the control. This situation is especially checked in the `Click_miCommentFont()` method.

```
private void Click_miCommentFont (object sender, EventArgs e)
{
    … …
    long idCmntCtrl = cmntPressed .ParentID;
    GraphicalObject grobj;
    for (int i = mover .Count - 1; i >= 0; i--)
    {
        grobj = mover [i] .Source;
        if (grobj is CommentedControl && grobj .ID == idCmntCtrl) {
            CommentedControl cc = grobj as CommentedControl;
            cc .CommentFont = dlg .Font;
            cc .CommentEnforcedRelocation (mover);
```

The `CommentedControl.CommentEnforcedRelocation()` method checks the positions of comment and control; if the comment is totally closed by the control, then the comment is slightly moved to the side.

I mentioned the situations, when the fonts' change can lead to such result in elements' positioning that will require some of their relocations by the user. The standard reaction from an experienced programmer would be to think about such possibilities and to add a small bit of "improvement" from his own. My advice: don't do it. Such even small addition is against the full users' control of the applciations. Users are not fool or dull; you are one of them, and you are as smart, as anyone else. Users do with the applications exactly what they want to do. If you think that you are capable of organizing the good view of any program, with which you are working, then the users of your applications are capable of managing the programs, which you develope for them. If you think that they are too dull to do such a thing, then look into the mirror: you are one of users. We are all users; there are no exceptions.

**Rule 4. All the parameters must be saved and restored.**

There are many objects in the **Form_PersonalData.cs**; parameters of any element can be changed, so all these parameters must be saved until the next opening of the form. For saving / restoring of the viewing parameters, I prefer to use Registry, though exactly the same thing can be organized via file.

```
private void SaveInfoToRegistry ()
{
    … …
    regkey .SetValue (nameSize, new string [] {ClientSize .Width .ToString (),
                                               ClientSize .Height .ToString (),
                                               bShowAngle .ToString () },
                                RegistryValueKind .MultiString);
    groupData .IntoRegistry (regkey, "Data");
    info .IntoRegistry (regkey, "Info");
```

Saving / restoring via Registry and file are demonstrated in a lot of classes, used in this application; look at the classes in the **Form_Village.cs**, or in the **Form_PlotsVariety.cs**. Restoring of an `ElasticGroup` object is as simple, as its saving; the only difference is that for restoring there is an additional parameter - an array of controls for this group. In this array, controls must be provided exactly in the same order, as they were used for group's construction.

When everything works as smoothly, as you can see in the **Form_PersonalData.cs**, you may, at some moment, ask yourself a question like "How something like this can be designed without moveability of all the elements?" Well, it can be in one way or another, but even the best results would look, let's say, "very awkward" in comparison with the demonstrated form. Without any doubts, if the program is designed on non-moveable elements, then users would have to work with the provided variant and would have to be happy with it. Certainly, it would be better, if such users don't know anything about the user-driven applications. But what if the users already know that there is an easy way to design the same applications, but with the full control transferred to them – users, which will provide them with a lot of new possibilities? What they will think then, if in parallel with their knowledge about the user-driven applications, they would continue to receive the new versions of the old style programs with the motto "You have to like whatever we are giving you". It's an interesting situation to think about.

We have looked at two different programs, dealing with some data. The first sample – *Years selection* – demonstrated a well-known task, which was implemented many times in different applications. Implementation in the form of a user-driven application didn't add even a bit to its functionality, but made the program much more flexible and allowed to get rid of several problems in its design.

The second sample – *Personal data* – also has old style analogues, but the difference now is not only in the increase of flexibility. The level of flexibility is so high and with the control, passed to the users, they receive so many opportunities, that the whole work becomes absolutely different.

The next sample – *Data world* – is a kind of program, in which the unlimited flexibility becomes not the feature to speak about, but a mandatory thing, without which this application simply can't exist. This is the sample from the class of instruments, about which I wrote in the introduction.

## Data world

"In the beginning there was Chaos." I remembered this phrase, when started to listen to the new task, but I have to admit that such description would consist a bit of exaggeration, as it happened some years after the World's creation. Then another well-known phrase waked up in my mind, while I continued listening: "Go there, don't know where, bring me that, don't know what." In Russian folklore, these are the words, in which some despot ruler formulates the task to one of his subjects, of whom he wants to get rid of. I don't think that that was in mind of my old friend and the head of our department of mathematical modelling, but I remembered that famous phrase from nearly the first moment of explanation, and it only sounded louder and more clearly with every additional detail.

Physicists deal with different problems. For each particular problem, there is a specific set of constants, data, restrictions, and so on. For some of the tasks these sets can be similar or even identical to high extent, for others they can be absolutely different. There are also different stages of any scientific research, so if at the beginning some data can be used without any restrictions in order to find some kind of solution, later it must be used within strict limits, which were calculated and defined on the previous stages. So my colleagues needed some kind of a program, which would allow to combine the arbitrary sets of data of different origin (numbers, strings, arrays) and organize any piece of this data in such a way that it would be prepared in the best form for using in other programs at later stages in their research work. Did you understand anything? "Bring me that, don't know what." I can ensure you that there were nearly no more exact details, than you can find in the previous sentences. Oh, yes, there was one addition. It would be nice to have different ways of functions' definitions: by the sets of precalculated numbers; or typed in as (x, y) pairs, or simply by rearranging its graphical presentation with a mouse. They wanted all of these things working together! Can you do it? I like such programming tasks!

Not every program has to be turned into user-driven. If a task has very strict and definitely unchangeable system of rules, describing the relative positions of all the parts, then you can make the whole object moveable, but not the parts of it. The good sample of such task would be a chess board; I don't think that it would be a good idea, if users would be allowed to change the format of the board to anything different from 8 * 8. On the other end of scale, there are tasks, for which nothing can be determined beforehand. As I demonstrated in two previous samples, there can be simple or complex situations, in which the objects' moveability is very helpful. Now we have come to a task, in which such moveability is not only a helpful feature; I don't think that without it there would be any even relatively good solution. When the task is nondeterministic, it can't be designed with a predetermined interface.

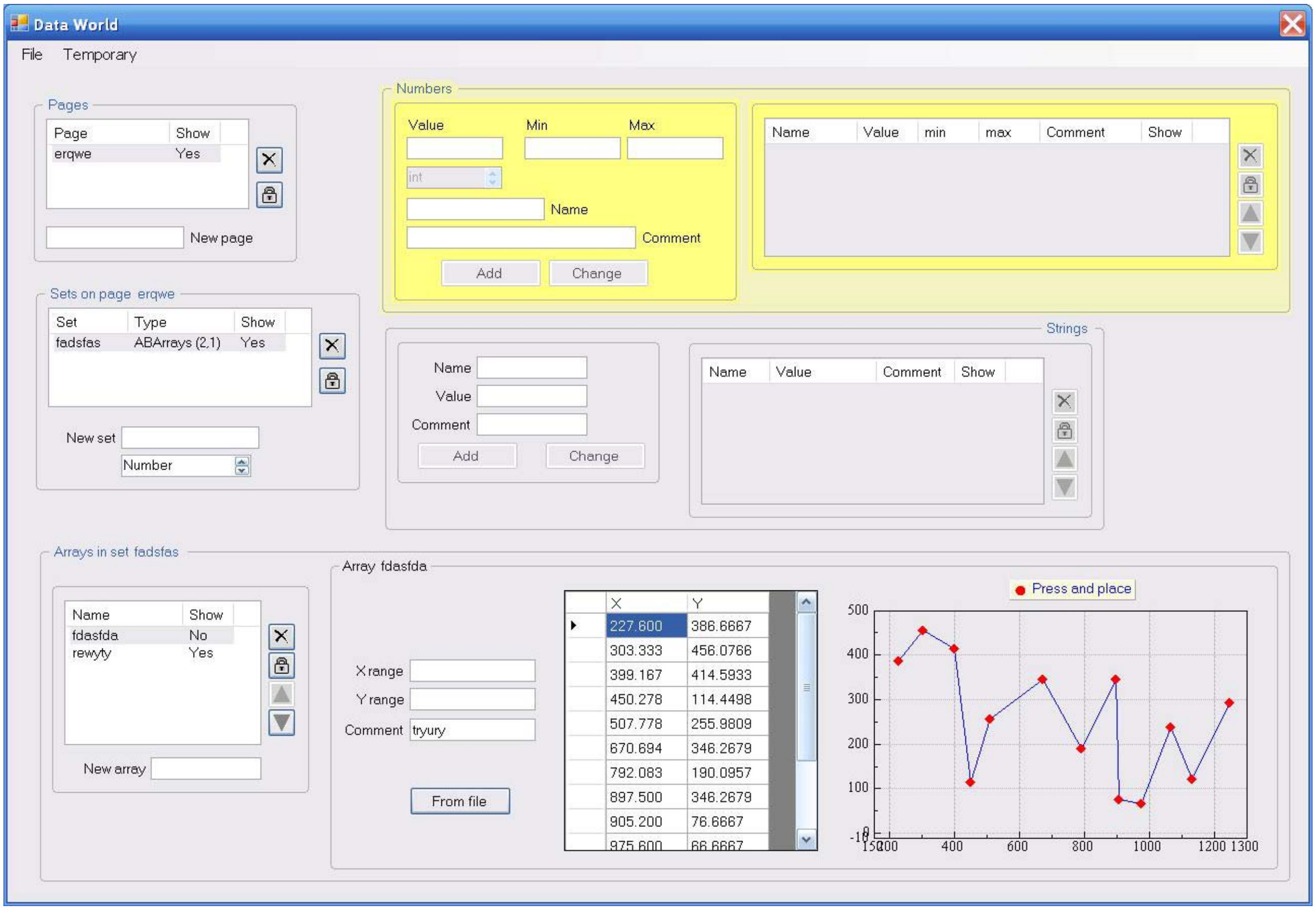

**Fig.4** Data world

**Figure 4** shows the view of the **Form_DataWorld.cs**. It is constructed mostly of the `ElasticGroup` objects, but an appearance of the plotting area with interesting features of editing the plot, required some extra work. I am sure that in the coming months this application will become even more complex, as one type of possible data is still missing. I want at least this version of the program to be used for some time; this will give the very important and highly needed information for its further development.

Several words of description; otherwise it would be impossible to understand anything at all in this task. The whole bulk of data, which the scientists are going to prepare with this program, is divided into a set of *Pages*. (They call them *Pages*; it has nothing to do with any book.) To organize a new page, it's enough to type a unique name and press Enter. Each *page* can include any number of *sets*; each set has a name and a type. The set's type defines the type of its elements: numbers, strings or arrays. When a set is selected in the list, then the appropriate group becomes enabled to define the new elements of this set. New numbers and strings are defined via a set of controls; this approach is standard enough. Y(x) functions can be defined in different ways: there is a `DataGridView` object, in which the values can be edited; there is also a graphical presentation of the function. New values can be added to the function by moving the red dot from the "nest" outside into the plotting area; usually it is the best way to start adding values to the new function. Function can be changed by pressing its line with a mouse; the new dot appears at this spot and can be moved to the needed location. Any existing dot can be relocated with a mouse or deleted via its context menu. The scales of the plotting area can be changed via their tuning forms; more about this tuning will be in further discussion in the chapter *Scientific and engineering applications*. I don't think that I need to go into further details; scientists use very special programs from time to time. This program is mentioned in this article only as a sample of an application, which, from my point of view, can be designed only as a user-driven application.

The program is much more complicated, than the previous sample, but it is based on the same rules and very similar in work. Nobody knows beforehand, which group of data will be of greater importance (and of greater need) at one moment or another, so everything is moveable and resizable. The groups for different types of data can be hidden from view, thus giving an extra space for more valuable information. Certainly, there are wide possibilities for tuning all parts of this form.

This is the third sample in the group Data. There are some similarities in these tasks and in my solutions of these tasks; that was the main reason of uniting them into a *Data* group of samples. There are also features that are common not only to these particular programs, but to a wide variety of applications; I would like to emphasize some of them.

These three applications, dealing with data, represent the programs of different levels of complexity. We started with a really simple program, based on 5 controls; then we moved to a more complicated one, which uses 23 controls. The last application, with 43 controls in the current version, is going to have even more in the nearest future. Complexity of these tasks increases not as a linear function of the number of used controls, but much faster.

With the first task (*Years selection*) there is a chance that users would agree even with the fixed design, because of the simplicity of the task. With the second one (*Personal data*), I doubt that such a trick will work, because even with the best proposed, but fixed design not more than 10 percent of users would agree. With so many possible variations in design, there are no chances for any one of them to satisfy the majority of users. With the third task, I wouldn't even think about such a possibility. When I work on one or another very complicated program for physicists, I often think about the possible solution in case there would be no moveable elements. In this case, I simply don't see such a solution.

A lot of people like to use `TabControl`; it's a good control for non-related or weakly related big pieces of information, but if you have to jump all the time between the tab pages, it becomes a nightmare. With the increasing amount of data, a single screen becomes too small to visualize it; more and more people begin to work with two or three screens. If the program is fixed on showing data with the help of `TabControl`, then these additional screens will not help; if the program is designed as a user-driven application, it allows the most efficient way of using any available screen space, regardless of the number of screens.

I am trying to underline one very interesting and extremely important feature of user-driven applications: the benefits of turning to such type of design increase even faster than the complexity of the tasks.

At the beginning of this part of an article I formulated a very important rule for design of different forms: any form can be constructed of only three types of elements (controls, commented controls, and groups). This rule works perfectly in many cases; in all the cases, when it's enough to have controls and comments to them. Some problems start to occur, when you have to add different types of graphical objects into the world of controls.* While including a plot into the **Form_DataWorld.cs** (**fig. 4**), I used one solution to this problem, but the more promising way is implemented in the **Form_ElasticGroupParams.cs** (**fig. 3**). In that form, there is a small group, which is used for setting the background color and transparency; this group includes two controls and one graphical object. That group belongs to the `ArbitraryGroup` class; further improvement and development of this class is of special interest to me. The whole world of moveable / resizable objects is like a big new continent, which has to be explored.

The previous sample belongs to a group of very complicated applications, but it is so specific that very few people would understand some of the details and would like to go into such details. Let's take another sample of user-driven application. In absolute contrast to the previous sample, everyone is familiar with this task.

# Calculator

The *Calculator* application, if there are no aberrations of my memory, appeared with the very first version of the Windows system, so regardless of when you personally became familiar with this system, the *Calculator* was already there. Similar applications exist for all other operation systems. The main thing is that everyone is familiar with this application and can make his own opinion on the usefulness of turning it into a user-driven application without waiting for some authorized opinion.

Up till the last fall I had no idea that I had to design my own version of this program. I use the standard Calculator from Microsoft three or four times a year, when I need to divide two big numbers and too lazy at the moment to do it with a pen

* In the first article I showed that you can do everything with combination of different graphical objects; some problems occur, when you start adding controls to the graphical world. Now we switch to the mirror situation – some graphical objects have to be added to the world of controls – and similar problems start. These two types of elements don't want to coexist.

on a sheet of paper. On those rare occasions, when I have to use the standard *Calculator*, I have trouble with finding in it the needed numbers, because they are definitely not at the places, where I would put them as a designer. It's a minor problem; if you use this *Calculator* as often, as I do, you can spend several extra seconds on such a search and then forget about these problems at least until the next season. But on one autumn day I heard a complaint from one of the colleagues: having a poor vision, the colleague had big troubles with the *Calculator* program, because even the combined efforts of several specialists didn't reveal any way to increase the font, used by this application. Another mystery from Microsoft. So I sat down and developed a Calculator, which that colleague could use. Certainly, changing of the font was not the only thing that I allowed to do in this program; it was designed according with all the rules of the user-driven applications. **Figure 5** demonstrates a possible view of my *Calculator*.

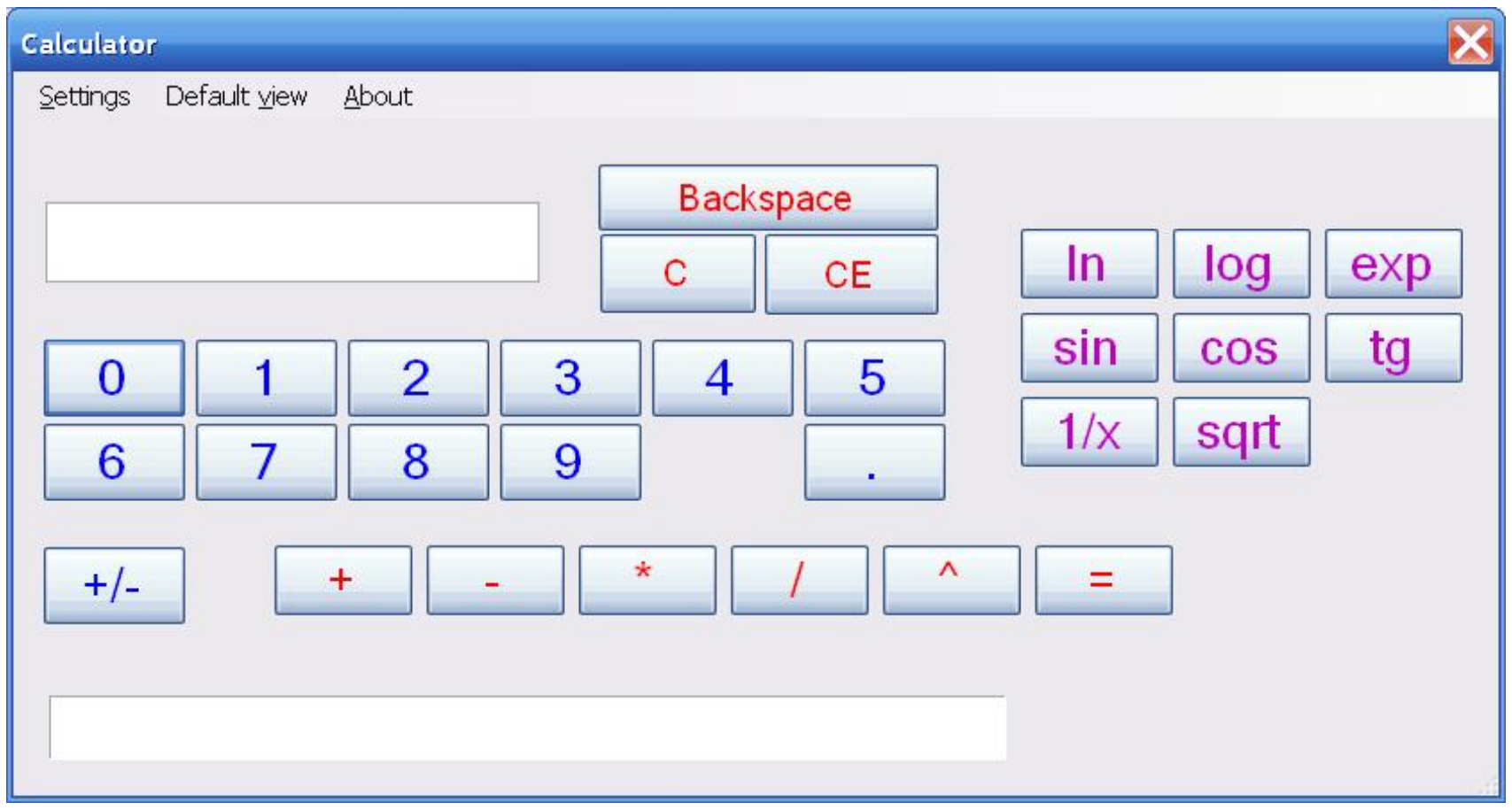

**Fig.5** Calculator

As usual, all the objects are moveable and resizable. In the first article, there is a special chapter about dealing with controls. You can find there the explanations for moving controls individually, in predefined groups, and in arbitrary groups; all three variants are used in my *Calculator*.

The three groups of controls are obvious: numbers, functions, and operations. On the figure, the buttons of these groups are shown with different colors. (From the point of mathematics, dot and sign are definitely not numbers, but they are used as a part of notation, while writing numbers, so in this program the two corresponding buttons are added to the group of numbers.)

- The controls can be positioned in any way you want; for specific positioning the individual movement of elements is used.
- Elements, belonging to three mentioned groups, can be visually united into a real group via the menu (for example, use *Settings – Frame digits*). All elements of a group can be moved synchronously. If you organize in such a way one of the predefined groups and other elements happened to be inside the boundaries of this group, then those extra elements are not included into the group. These three groups are the closed societies, though they are not secret and announce their names in the form of the group's title.
- Any group of objects can be united into a group by drawing a rectangle around them with a mouse. Free union of free elements. Elements of such group can belong to the different predefined groups or to the same one; there can be also those controls that do not belong to the predefined groups. Even if thus organized group consists of elements of one predefined group and includes all its members, it is still named *Arbitrary*.

When you have a lot of elements in the form, you can think out a lot of useful possibilities. You can add all those possibilities to your program and … the consequences depend on the personal view of each user. On one end of spectrum are those, who do not want to do anything and require from a program to do at any moment exactly what they are thinking about. (The computer has to understand my thoughts and decipher them correctly.) On another end are those, who want to have an absolute control of everything in a program. I sympathize more with the second group.

For three predefined groups of controls, there are some predefined variants of their relative positioning, which are set via the menu; for example, try *Settings – Digits – Standard positioning*. But these are only some of the possibilities among many others, as you have also the individual movements and arbitrary groups.

The same variety of choices exists for setting of other viewing parameters.

- The size of any element can be set individually.
- The size of any element from three groups can be used as a sample for all other elements of the same group, or even spread on the elements of another group.
- Color can be declared for all the elements of the predefined groups.
- Font can be declared for all the elements of the predefined groups. The same font can be spread on all the objects.

There are other possibilities, which can be easily added to already mentioned above, but I don't think that it is needed. It's so easy to give users an opportunity to change individually the font and color of each element. This is demonstrated in the next sample, but I am not sure that it is needed in the case of *Calculator*. If you want to have a chance of applying different colors to different numbers, you can easily do it by adding several lines of code; all the codes of this application are available.

I have demonstrated FOUR tasks with absolutely different purposes and at different levels of complexity. All these applications work under the same rules; only the level of parameters' tuning can vary, but even the tuning itself is organized in the similar way, so if the user knows that a program is developed as a user-driven application, he will have no problems with finding and using all the program's features and possibilities.

These four applications are designed on the basis of controls. There are other areas, in which the programs are more oriented on graphical objects. Let's look at such samples and see, if such change of the basic elements can bring anything new into the design of the user-driven applications.

There is also one more change with a step from the previous samples to the next one, and this change is much more important. Usually the world of controls is more determined than the world of graphical objects. (At least, it's my feeling when I compare these two big areas.) For example, if an application is supposed to deal with the personal data, then there is some description (a list) of data, which is going to be used. You design an application to work with this full set of data; user can select any subset of the whole amount, which he needs at one moment or another. For the *Calculator*, you can add several more buttons for additional functions, but the overall behaviour of such application is not going to change. Moveable / resizable elements change the way, in which people can work with such applications, but the number of needed controls is usually defined by the idea of the task. (There are some applications, in which the variable number of controls is determined at each moment by the current status of the task. Such applications are relatively rare in the world of controls and are more close to what you'll see further on.)

There are other areas, and these are mostly different areas of research work, where the limitations of the programs are mostly determined not by the tasks' requirements, but by the designers' abilities to solve those tasks. More correct to say that the limitations of such programs are determined *by the designers' abilities to solve the interface problems*. Researchers have the unlimited requirements; programmers often cut those requirements to a very limited area. The use of moveable objects gives a chance to move these limitations far aside and provide the researchers with much more flexible instruments for their work. Let's look at some samples.

# Functions' analyser. Scientific and engineering applications

For years the computers were used mostly as very powerful calculators; when the graphical displays appeared later, they were immediately used for better presentation of data and results in scientific / engineering applications. The history of graphical presentation of different functions includes several decades. All libraries on all types of computers provide the functions' drawing, but programmers continue to write new and newer methods of their own, because the existing standard methods are too general and often not the best for each particular case.

I started to work on the programs with the functions' presentation in pre-PC era; in computer history it is the analogue of the Middle Ages. The hardware and software changed throughout the years, the new programming languages were born, the new ideas in programming pushed a lot of people (and me also) into rethinking of some basic principles, but for many years I work mostly on scientific / engineering applications in different areas, and the core of all the applications in all these areas is still the same: "Draw me the function".

There are two milestones in design of scientific / engineering applications during the PC era. The first important moment happened, when the monitors' textual mode was abandoned and the programs began to work exclusively in graphical mode. Then the increasing processor speed allowed to use more complicated (and time consuming) calculating algorithms in parallel with real-time graphical presentation of results. For several years the hardware progress ignited the new ideas in overall design of scientific / engineering applications, making possible the things, which couldn't be implemented on slower computers.

Somewhere 10-12 years ago the design of scientific / engineering applications went into the period of stagnation. The new versions of the big and well-known programs are still distributed among the clients every year or two; managers of these projects continue to declare each version as the biggest achievement in the history of mankind, but all those words do not worth a paper, on which they are printed. If there is really anything new in those programs, then it is only in the algorithmic part (the core of each program), but not in overall design.

This is really a strange situation: the continuing progress in hardware; the C# language, which is much better than C++ for programmers' work on big and small scientific / engineering applications, and yet there is nothing new or helpful for users of these applications. Users don't care about the language or the programming environment that the designer used throughout his work; they care only about the result and what they get. And this is absolutely correct, because users must do their own work, and the only important thing for them is: in what way the new version of this big program is better for my work, than the previous version?

I worked on the big scientific / engineering applications for different areas, and it became obvious to me that the problem was not in some particular area, but it was a general problem. This general problem has its roots in the main designing idea: applications are designer-driven.

The users of scientific / engineering applications are often better (much better!) specialists in their particular area, than the designers of the programs, with which they try to solve some very complicated problems. Thus the better specialists have to work inside the range of understanding, enforced by lesser specialists. The only way for the users to get something new from the programs is to explain to the manufacture that they really need this and that features. If they would be successful in this, then maybe in a year or two they would see some additional parts in the application, but nobody knows beforehand, how those users' explanations would be interpreted by the designers.

You can find in many books the description of the adaptive interface as a solution to "user – designer" problems, because it gives user a chance to select the best solution personally for him. What is never written and carefully hidden is the fact that user gets a chance to select only between the choices that were predetermined by a designer according with his understanding of each situation and of his vision of the best solutions for users in each particular case. Adaptive interface can soften the problem, but not solve it. When I understood that the main problem of all the scientific / engineering applications is in their being absolutely designer-controlled, I also understood that this hurdle can't be crossed with the help of any known or will be thought out in the future form of adaptive interface. Only something different would allow to solve this main problem and to move forward. Then I came to the idea of the user-driven applications.

It was obvious to me that such applications have to pass the control over the screen elements to the users, thus the algorithm of moving the screen objects became crucial and had to be invented. When I thought out such algorithm, I immediately tried to apply it to the objects, which I was using all the time in all my programs – the scientific plots. Maybe it was a mistake and I had to begin with much simpler objects (in the way I demonstrated this algorithm in the first article), but an attempt to use this algorithm on complicated objects unveiled to me some basic features of the user-driven applications and helped to understand a lot of very important things for the whole area.

Plots, used in scientific / engineering applications, are not very simple objects. They have a lot of parameters for visualization and can be used in different situations. I work with the plots (classes) of my own design for years without nearly any changes at all. They were carefully checked, they worked in all the situations, and I didn't expect any problems at all. Initially I thought about turning the rectangular plotting areas into moveable and didn't want to change anything else, because everything worked fine. Such view on adding the moveability to the reliable and working classes of objects turned out to be absolutely wrong. My understanding of the situation improved step by step. Only the first of these steps was a voluntary one, as a result of thinking about the whole problem; all further steps I was enforced to do by the logic of the design.

**Step 1 – main plotting area**. I took one of my big applications with a lot of different plots, turned the main plotting areas into moveable / resizable, and began to play with this new program. It was a huge improvement of my program, but the more I worked with it, the stronger became the filling that something was incorrect. The logic of design was perfect for a fixed designer-driven application; even turning of a single part of it into moveable / resizable began to corrode the whole construction. The features, which were thought up to the tiny details and were used for years without any problems, didn't want to work smoothly with the new plots. The logic of moveable and still non-moveable parts began to conflict.

Plots have different scales; usually the scales are positioned near the plotting area; in the old programs with unmoveable parts, users could call the tuning form of a scale and type in some parameters (for example, the distance between the scale and the plotting area), which would change the scale's position. The moving / resizing of the plotting areas by a mouse change the positions and sizes of the related scales in an appropriate way, but the individual relocation of the scales was still possible only via the tuning forms, and it looked very awkward. If a plot with all its scales can be moved around, why the scales themselves can't be moved in the same easy way?

**Step 2 - scales**. Certainly, I designed this next change, but it wasn't simply a transformation of another graphical object into moveable. Plots and scales are the strongly linked objects with the type of "parent - children" relation. So now there were objects (plots), which could be moved synchronously with all their related parts, but at the same time those parts could be involved in individual movements. That required some type of identification for the objects, involved in moving / resizing; this identification system has to guarantee the correctness of synchronous and individual movements for the objects with any type of relations between the parts. It had to be not the new type of identification for each new class of objects, but the general solution for any classes, which will be designed in the future and involved in different types of movements. (This system of identification was described in the first article.)

**Step 3 - comments**. With the plotting areas and scales now moveable in any possible way, my attention turned to another huge problem, for which neither knew a good solution before: good positioning of the comments along the graphs, calculated by a program. In all the numerous scientific and engineering applications the plotting areas are better or worse positioned by the designers; all these plots are unmoveable, so the designers simply decide about their positions, and that's all. The results, which are shown in those areas, often need some comments, but the exact lines' positions are calculated throughout the work of applications, so there is no way to determine beforehand the good position for these comments. When an application produces some results in the form of the plots, then users are given a chance to position some comments next to these lines by, for example, typing some positioning coefficients in the tuning form. In such a way it was done before, but now it looks very strange if the plots and scales can be moved easily around the screen, but the comments to these objects are positioned and changed in some archaic way. The comments have to be moved exactly in the same easy way, as plotting areas and scales – by a mouse. In addition, the comments need to be not only moved forward but also rotated for better placement along the arbitrary lines.

I think that now it is obvious in which direction the scientific applications started to change after a single element – plotting areas – was turned into moveable / resizable. There is a law that the reliability of the whole system can't be higher than this characteristic for the lesser reliable part. The similar rule, translated into the world of programming, declares that the flexibility of the system can't be higher than this characteristic for any part.

**Step 4 - controls**. Even if you turn all the parts of the plots (areas, scales, and comments) into moveable and resizable, but leave in the same form (dialogue) anything unmovable, then the users would always bump on this hillock, even on a small one. The inner area of applications is populated both with the graphical objects and controls, so controls and groups of controls must be turned into moveable / resizable. I made this change, received the new type of scientific application, and that was the new paradigm – *user-driven applications*.

**Step 5 – tuning forms**. But that was definitely not the end of the road. Everything became moveable / resizable in the main form of an application, and it looked fine. The positions and sizes of all the elements are now easily changed with a mouse, so each user can rearrange the view of an application to whatever he wants at any moment. But positions and sizes are not the only parameters, on which the view depends. All the plots, scales, and comments have their parameters of visualization; these parameters can be changed in the auxiliary tuning forms. After playing with the improved plots for some time, I decided to change some of the visual parameters, opened one of the tuning forms and tried to move the elements there. It was the most natural thing to do after all those moving / resizing in the main form, but it didn't work: all the tuning forms were still old-fashioned. I began to redesign all the tuning forms on the basis of moveable / resizable elements.

It wasn't some kind of whim; there was one more thing, which required redesigning of those tuning forms according with the rules of user-driven applications. The tuning forms for all kinds of scales include a sample of a number or a text, which shows the positioning of all numbers (texts) along the scale. This sample is moveable; its movement / rotation is copied by all the numbers (texts) along the scale. When you have the moveable element inside the tuning form, then the mentioned conflict between the moveable and non-moveable elements immediately jumps into this form, and you have to redesign it.

The next comment came from my colleagues, who began to use more and more of the user-driven applications in parallel with other programs, with which they work for years. The colleagues caught themselves on trying to move the objects in those programs exactly in the same way, as they move them in my applications. Certainly, it doesn't work (all other programs are designed in an old way), but each time it happens like a mini shock to the users: why this object doesn't want to move, when every other is moveable? It takes some part of a second to understand the reason and to remember that this is a different type of program, in which everything is still fixed…

Now I found myself in the similar situation. I was working for many hours on some part of the program, accompanying this article. I was checking different parts in order to find any possible mistakes; I was switching from one form to another, and while I was in the main form, I tried to move it simply by some inner point, as I move any graphical object. Certainly, the form didn't move in such a way, and my first thought was about some mistake, which I had to find. The next instant I remembered that that was not one of my graphical objects, but the form itself, which could be moved only by the caption bar. After 20 years of working with Windows, I tried to move a window by its inner point! When it happened again, I started to think that this is not the problem of ageing, but the demonstration of the basic rule for the user-driven applications. You quickly got used to the moveable / resizable objects and expect that all the objects in all the applications have to work in such a way, because it is the most natural way of dealing with all of them.

Moveable / resizable objects do not want to coexist with the fixed objects in any way. **If there are moveable objects in an application, they will insist, require, demand that all other objects around were turned into moveable / resizable.** This requirement is not limited by the form, in which these moveable objects are placed, but spreads on all the related forms, on all the forms of the same application, on other applications… The expanding universe of the moveable objects. (Only in this case I know exactly, how the Big Bang happened.)

**Step 6 - unexpected**. Whatever is mentioned several lines before is definitely not the new thing for me: colleagues mentioned with laugh their attempts to move windows by the inner points (not by their title bars!); several times I had tried to do the same. And when I have already finished this article and was rereading the text once more before publishing it on the web, the idea came into my head that this is not a problem at all. If the logic of applications, filled with moveable objects, demands the moving of the forms by any inner point, then it can be done in seconds by adding several primitive lines. It doesn't even require the use of mover or any special class; simply change the form's location synchronously with the mouse move, if the form was pressed by a mouse at any inner point. To do such a thing, you need a very simple addition to the same three mouse events: `MouseDown`, `MouseMove`, and `MouseUp`. The code is so primitive!

When you press the left button anywhere in the form and the mover doesn't catch any object, then the shift from the mouse to the form's location is calculated; the flag is also set to inform that the form is in move.

```
private void OnMouseDown (object sender, MouseEventArgs e)
{
    ptMouse_Down = e .Location;
    if (!mover .Catch (e .Location, e .Button, bShowAngle)) {
        if (e .Button == MouseButtons .Left) {
            bFormInMove = true;
            sizeMouseShift = new Size (PointToScreen (ptMouse_Down).X - Location.X,
                                      PointToScreen (ptMouse_Down) .Y - Location .Y);
```

If the mouse is moved without moving any object and the flag shows that the form is in move, then that calculated shift is used to define the form's position

```
private void OnMouseMove (object sender, MouseEventArgs e)
{
    if (mover .Move (e .Location)) {
        … …
    }
    else
    {
        if (bFormInMove) {
            Location = PointToScreen (e .Location) - sizeMouseShift;
```

I added this feature only to two forms in this application: **Form_Function.cs** and **Form_PlotsVariety.cs**. Both forms include a lot of moveable objects; it would be absolutely natural, if the forms themselves can be moved by any point, when the same technique works with all their inner elements. You can try this thing on any of your forms, or on other forms in this demo application and decide for yourself, if such an addition to moving of the standard windows can be useful. For 20 and plus years I was moving windows only by the caption bar; then the work with the user-driven applications pushed me into thinking that there can be a very useful addition even for such a well known thing. That's how the user-driven applications affect all the things around.

The user-driven applications were not born as a pure scientific idea, though it wouldn't lose a bit of its importance if it would happen this way. Such programs were born because I designed an algorithm of turning an arbitrary screen object into moveable / resizable and began to build applications on the basis of such elements. Though the user-driven applications turned out to be of high value in many cases, the area of scientific / engineering programs is the one, where the new ideas not only improve the applications, but turn them into the most flexible and effective instruments of research. To understand the important details of such transformation, let's look on one program from this area. Usually, it's very difficult to explain to non-specialists the idea of some specific scientific application, but I am going to demonstrate a program, which anyone will understand.

Suppose that you want to look at some function for yourself, or you teach some course, in which you need to demonstrate different functions, to speak about them, to compare, to attract the attention to function's behaviour in one or another area of existence. In any of these cases, you need an instrument to define the functions, to show them in different ways, and to store them for later use. For all these purposes we have **Form_Functions.cs** (**figure 6**).

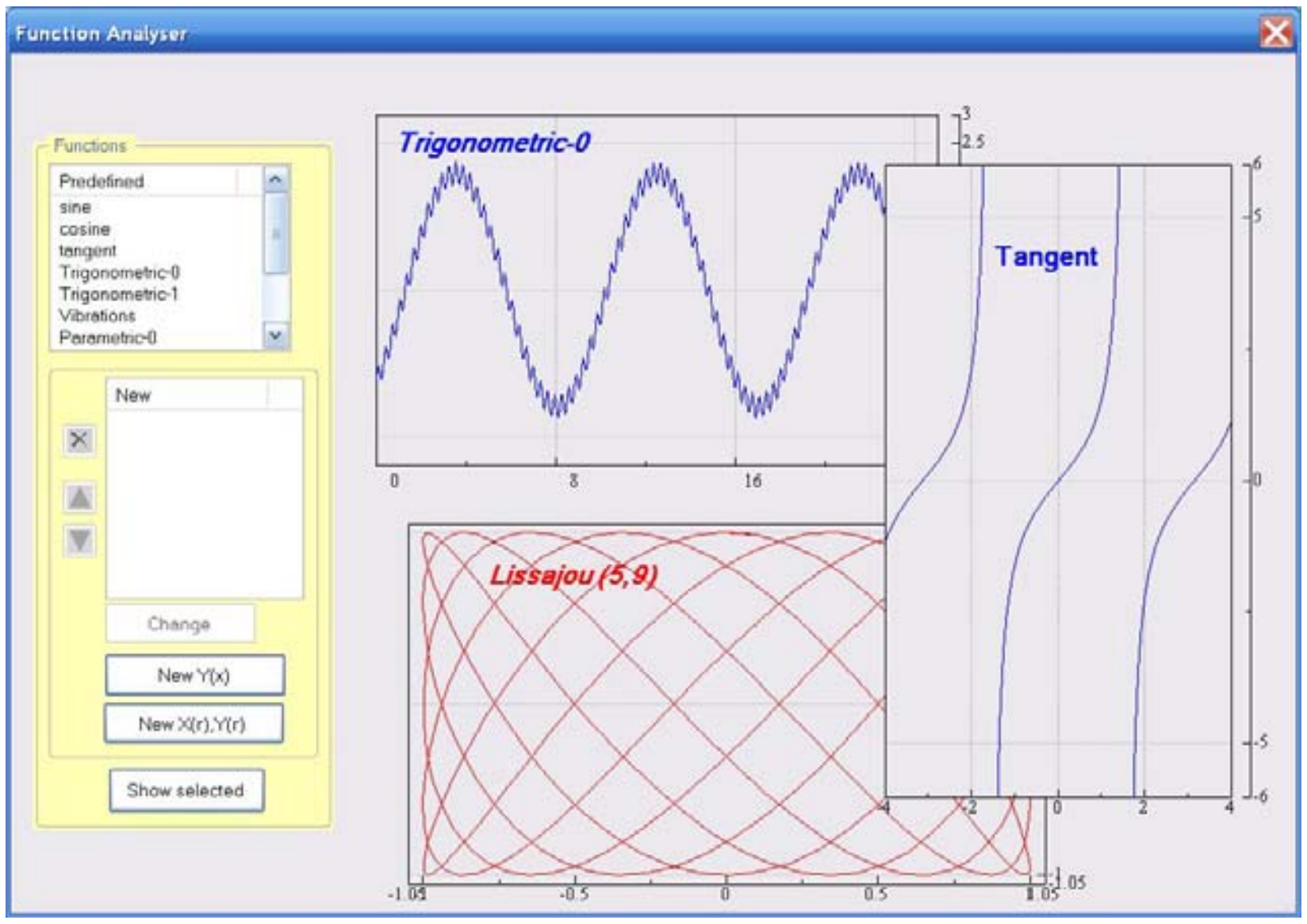

**Fig.6** Functions' analyser

In addition to this form, there are two others, in which you can define functions of the Y(x) or {X(r), Y(r)} types. The names of these new functions fill one list in the group at this figure; there is also another list with the names of several predefined functions. These predefined functions are included into the application only to allow trying this form just from the beginning without going into the process of new functions definition. Certainly, for your own work the predefined functions are not needed; this list can be closed with one command in the context menu of the group.

This is a classical user-driven application. As a designer, I constructed an instrument, which allows to look at the functions. In those two additional forms (they are called by clicking two special buttons) you declare any function you need; my instrument only checks the correctness of the function's expression, but nothing else. I am not supposed to put any restrictions on the number of functions or their visualization. You select any set of functions to be shown in any area; you select the position, sizes, and other parameters for areas' visualization. I have to provide the most flexible instrument, so that you would not feel any limitations from my side on whatever you will decide to do today, tomorrow or a couple of years later.

Do you remember the process of solving some really complicated problems? I hope that some readers still do it in the same way; others might have some memories about it. If you have to solve some equations or design some mechanism, you take a sheet of paper and begin to put down the initial equations (or the first sketch of a mechanism). When there is no more space on the first sheet of paper, you take another one. You put the filled sheets somewhere on the table (or desk), they can be accurately piled, or thrown around in a mess, at one moment or another some sheets can be placed side by side to compare the previous results, or a sheet can be taken from underneath and put on top, because it contains some important information.

Nobody is telling you, how to place all those sheets, but each person organizes them exactly in such a way, which is the best personally for him. Millions of people use this procedure for centuries, a lot of people do it in similar, but not in exactly the same way. There are no regulations of this process; each researcher determines the procedure, which is the best for him and his work. The results of person's work differ a lot and depend on many things (brains, talent, education, circumstances…), but each researcher or engineer decides about the best procedure personally for him.

I have designed several applications of such type for my colleagues in our department and I like to watch from time to time, how they use those programs to solve some really complicated tasks. They take some part of the results from their experimental work and organize it into one plot, then another. The plots on the screen grow in number; the unneeded plots are deleted; the new plots are organized. They are placed side by side for comparison, or they can cover each other. At one moment somebody will take one from underneath and put it next to the new plot, because he remembered something and had an idea, how their comparison can explain something strange in behaviour of the experimental material. They work with these plots in the same way, as they work for decades with the sheets of paper.

You can do exactly the same with the **Form_Functions.cs**: if you want to show and discuss some function – you type in its text, add any number of comments and place them, where you want them to be. If you want to show several functions in one area – select these functions in the list(s) and click *Show selected*. You want the different areas of the same function to be shown in separate areas – do it this way. I prepared an instrument to define the functions and look at them; you, as a user, decide about the way to show them. Whatever is prepared, can be saved and later restored.

Plots and scales have their tuning forms, which can be called via context menus or with the double click. All these auxiliary forms are designed as user-driven. From time to time I change the default view of these forms (when I think out some better class for their design), but they are always user-driven forms, which can be rearranged by any user in the way he personally prefers.

The moving / resizing of the plots, which is demonstrated in the **Form_Functions.cs**, are not the only way, in which the new technique can be used in the scientific applications. The area of such programs demands the use of big and small moveable objects in every corner, in every part, but … The moveable elements in the currently used programs are so rare. Why? The total stagnation of minds. (Mine is not an exception; it took me too long to understand the problem and begin working on it.)

Here is one more sample, where the addition of moveable elements changed the whole process of analysis. This sample is not included into accompanying Demo application, but here is a picture and several words of explanation.

The areas of science put in front of the programmers the same tasks again and again until the new programming ideas would produce a really elegant solution. A lot of purely numerical problems were solved years ago, but the combinations of calculations and manual adjusting of the algorithms usually waits until some moment, when the new development in interface would allow to solve it easily. One of such tasks is the data refinement. A lot of researchers receive the needed data from the analog-digital converters. This data is going to be analysed by a well known mathematical methods, but the data is often obtained together with an additional noise, and the well known programs for calculations go crazy, while trying to solve the equations with such input values. The data has to be previously refined by a researcher, and that's where the moveability of different elements is of high demand.

**Figure 7** demonstrates one form from the **DataRefinement** application. There is a huge amount of data, received from experiments. Before using this data in the well known methods, a researcher has to look through and refine it. The system allows to cut the input data into segments and to select any part of input data; moveable green sliders make this task easy. Each (x, y) input value is shown on the screen as a small circle; any circle can be moved up or down, thus making the curve smooth by reducing the noise; then the curve on the selected segment is substituted by one of the well-known simple functions. The coefficients for such function are calculated by a special algorithm; such function gives a good approximation, and it is much better for the next step of analysis, than the set of dots.

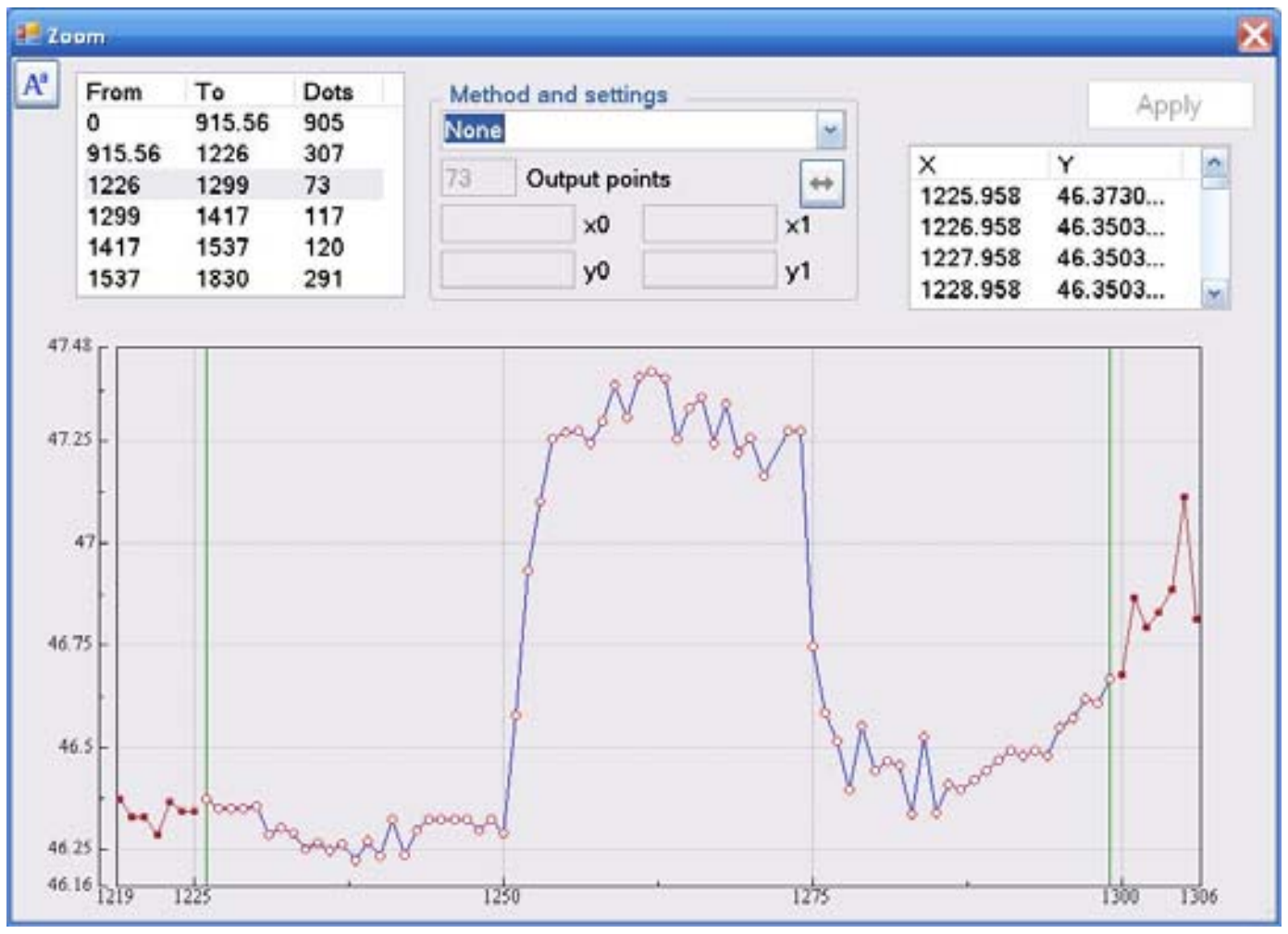

**Fig.7** Data adjustment in the **DataRefinment** application

The whole form is organized according with all the rules of user-driven applications, so the moveability of the screen objects is used on all the levels. I have no doubt that something similar can be developed without moveable elements, but such variant will look so awkward in comparison and so inconvenient in use. Colleagues were thinking about such an application for years, but they never developed it, because they understood, how ineffective it would be (retyping new values again and again in some kind of a table…). It took me several days to understand all the requirements and to give them this version, which works according with all their requests and expectations.

There is a wide variety of scientific and engineering applications with different requirements from their users. I have demonstrated the samples with the plots, which use previously calculated data; in other programs users can manually change the plots or even organize the new plots in such a way. Applications become not only the tools of visualization, but the instruments of research. How good are these instruments for the research work, depends on the developers' skills, but the main idea and the basic rules of design for such applications are changing.

I am designing very complicated scientific / engineering applications for many years; some of these programs are sold all round the world. The design of any new system always started with the discussions on all the details; the coding would never start until the specification would be written. For years I had no doubt that that was the correct and the only possible way of the systems' design. I wasn't the only one to think in such a way. Take any book on the theory of systems' design; as a sample of the worst scenario there would be the description of the work, which started before everything was discussed along and across and which started without the previously prepared specification on all the big and small parts of it. With the conclusion: NEVER do it in such a way; this will be a disaster.

My belief in the correctness of those well-known rules began to deteriorate four years ago. Some time after it the strong construction of those old rules crashed into pieces, and I began to understand the realities and rules of the user-driven applications.

Now I design not lesser complicated programs than before, but under absolutely different rules. There is only one layer, which still requires the detailed specification: the layer between the input data and the program. I have to discuss the details of input data; otherwise my program will never read it correctly. But nothing else is specified; after the program gets the data, users can do, whatever they want. The discussion of the possible changes often continues in parallel with the application's use and further development. It's against all the known rules, but not for user-driven applications. It works fine and provides much more possibilities for the users!

As I mentioned, there are some discussions at the beginning of any new project, but in details only the data format has to be specified. Not only programmers, but users also have to understand the realities of the new world. This depends on each person, but when you work with the talented scientists, it happens quickly enough. They get the new ideas and quickly understand, what new possibilities it opens in their research work. It would be impossible to give them now anything designed in the old style. It's like giving somebody an application, which works only in DOS. You can try and see for yourself, if it will be accepted or not.

The design of the scientific / engineering applications on the basis of moveable / resizable elements will require a lot of thinking about the principles of their construction. I formulated and explained four basic rules of the user-driven applications and tried to show, how those rules are transformed into many tiny details, which can make the difference between excellent programs and the bad ones. As the area of scientific / engineering applications is very big, there can be a significant difference between the requirements to such programs. Also the programmers will bring into this process their personalities, so I expect to see the wide variety of approaches. Especially, because it is the new universe without any detailed specifications. I underline again and again that USERS receive the full control over such applications, but there can be a lot of differences on how this control is passed and how easy it would be for users to do one thing or another.

Just one sample on one item only. For some time I thought that to achieve the best result, a designer has to turn any element into moveable, so the main problem was in organizing a cover for an arbitrary object. Certainly, it has to be done, but it turned out that it is not all, what is needed. When there are a lot of different moveable objects on the screen, users would like to have an easy control of their moveability level. In the **Form_Functions.cs**, the moveability of different elements can be regulated via several context menus:

- Elements of the group can be fixed / unfixed via the group's menu.
- Moveability of the scales, belonging to any plot, can be changed via the plot's menu.
- Any comment can be fixed or unfixed via its own menu.

Even the implementation of these things is organized in different ways. My colleagues, who use different user-driven applications with a lot of plotting, told me that they don't need the individual change of moveability for each scale: either all the scales of the same plot are fixed in relation to the plot or unfixed. Each scale has its own flag of moveability, so it's not a problem to organize the individual switch of moveability for each scale, but I was told that that was not needed.

Different approach to the comments' moveability: each one of them can be fixed / unfixed on the individual basis, but there is no line (command) in any context menu, which will allow you to set this feature for any group of comments. This is the result of my own uncertainty about this problem. If I'll include into some context menu the commands (lines) to fix / unfix a group of comments, then there must be a short, but obvious explanation, to which group of comments this command is going to be applied. The comments can be linked either with the main plotting area, or with any of its scales; any comment can be moved to any screen position, so it is impossible to decide about the exact comment's link (with area or with scale) simply from the comment's view or position. Because of this visible uncertainty, even the good explanation by the command's text is not enough. There can be different variants of performing such a command on different groups of comments: fix / unfix all that belong to the main plotting area, or to a single scale, or to all the scales of the plot, or do it for all the comments regardless of their link… There are many different variants, but in real applications I never saw too many comments, related to the scale or a plot. Thinking about all these things, I didn't add any commands to change the moveability of a whole group of comments; it is organized only on the individual basis. It's the situation for today; maybe I'll come to some different (and good) solution in a week or a month.

I am not trying to describe in details how the user-driven applications have to be designed; it would be absolutely against the idea of such applications. There are some basic rules, which I began to understand while designing more and more applications of such type; I'm sure that anyone, designing the user-driven applications, will come to the same conclusions. But outside these rules the user-driven applications are a vast unexplored area without history and habits.

The area of scientific / engineering applications has one BIG problem, which influenced for decades the design of needed applications and their use: the research work, which is unbound by its nature, has to be squeezed into the limits of the designed programs. This is the cause of the endless quarrels between the designers and the users of such applications. The switch to the user-driven applications eliminates this problem and turns the programs of this area into the real research instruments. I see it more and more, because with each new application, developed in our department of mathematical modelling, the researchers simply accept the new level of a program as a basic one and require the new things, which were looked at as something fantastic even several months ago. It is obvious that with the introduction of the user-driven applications the stagnation in the area of scientific / engineering applications is over and we receive a huge step in their development. Scientific and engineering programs often are very special and the majority of them have only several users. To develop a big and very sophisticated application for several people only is always a problem. In addition, such programs need frequent changes in interface; otherwise they become unsuitable very quickly, because the continuation of the research work constantly changes the requirements to such programs. Because of the whole knot of related problems, the needed applications are often not designed, as it is impossible to do under the ideas of the fixed design. The switch to the user-driven programs allows to solve a lot of problems in design of the very needed instruments for research work.

There is one absolutely unexpected result in switching to the user-driven applications. It's a shame and a disgrace to our profession that companies never put the names of the main developers somewhere in the *About* form; they put the name of the company, but not the personal names. However, the views of the big sophisticated applications often retain a signature of their main developer, because an experienced person usually design the forms in the same (or similar) way, in which he likes to work for years. I was told by the users of one of such programs that they saw immediately, what was the result of my work, because they were familiar with my style from other applications. For user-driven programs it would be impossible! You can't imagine how users change the views of applications, when they get a chance to do it. Now, when I am asked to look at an application to discuss with my colleagues the questions of its further development, I have always to start with the question on the name of this particular program, because it's often impossible to recognize an application, on which I continue to work. It's an unimaginable situation for the standard applications, but often happens with the user-driven programs. Never mind, the most important thing is that these applications work and look exactly like users want them to be!

# Variety of plots

The area of scientific / engineering program is the one, which demonstrates the advantages of user-driven applications at their best; there are some other areas, for example, data visualization, which can show similar results.

The world of data is infinitive. There are many forms of data visualization; there are systems (libraries), which allow to show data in different ways. Throughout all my professional life I worked on different scientific / engineering problems, which use mostly the standard plots of the Y(x) type or something similar; other types of plots rarely attracted my interest. But with the invention of an algorithm for turning objects into moveable / resizable, I was especially looking for unfamiliar objects to check the algorithm with the unusual forms, or in strange situations. The variety of plots, widely used in other areas, turned out to be a good range for my ideas.

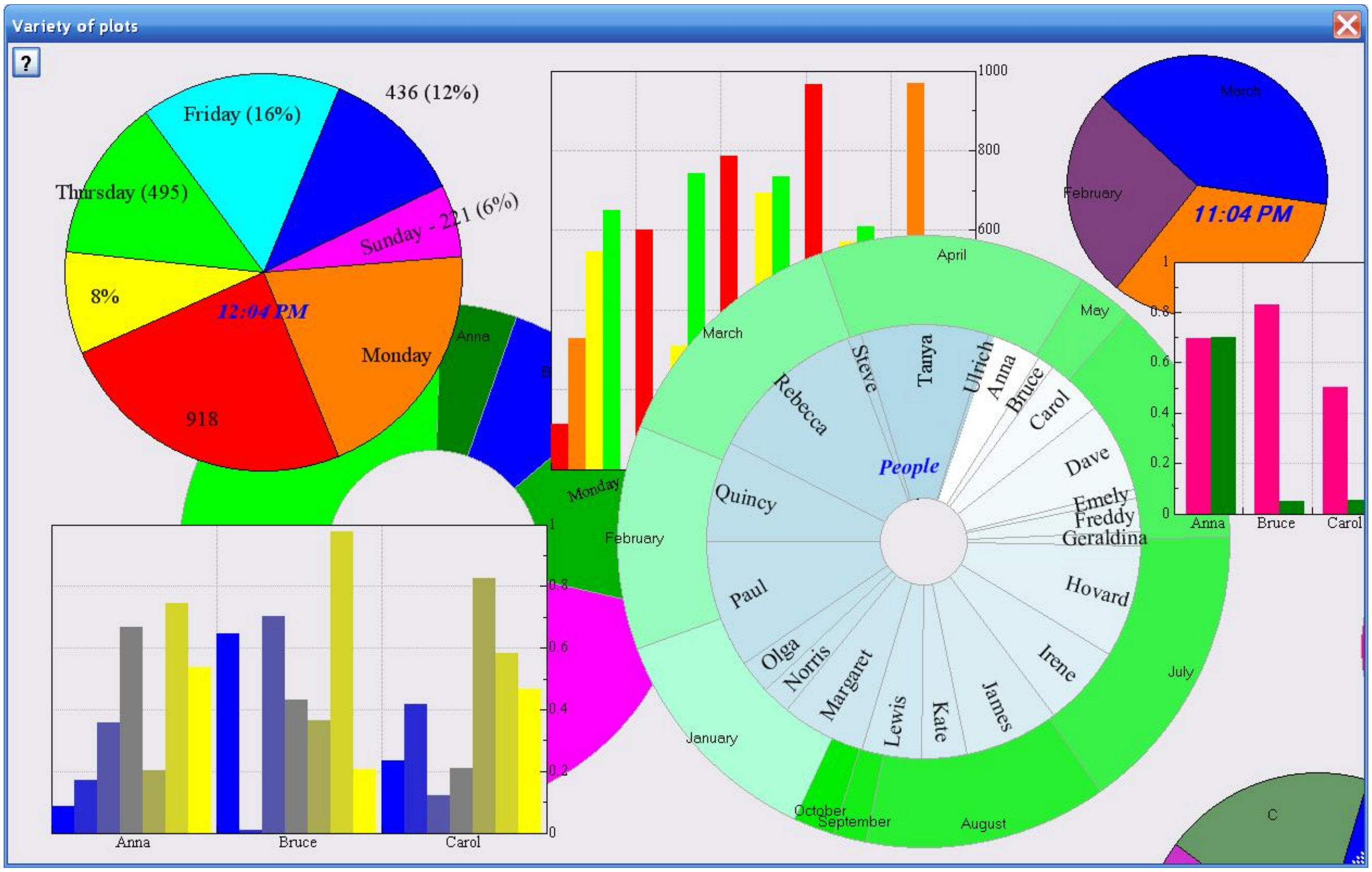

**Fig.8 Form_PlotsVariety.cs**

**Figure 8** demonstrates the **Form_PlotsVariety.cs** with several drawn objects. Several means the number of big, complex objects, like a bar chart or a pie chart. Each of those objects consists of a variable number of constituents, which belong to the different classes. All these objects are involved in individual and synchronous movements. Each object has its own visual parameters, which can be set on an individual basis, or a set of objects – siblings can get the same parameters for visualization. The variety of variants is provided by a designer; the way to use all these things is decided by the user and only by the user. There are certainly no limitations from the designer on the users' work.

All different types of plots, which you can see here, are designed under the same ideas, which were demonstrated with the scientific plots in the previous chapter. Each complex object is moveable by any inner point and resizable by any border point. In a set of rings, each ring is resizable both by its inner and outer borders; also any number of rings can be added to such object.

A bar chart (`BarChart` class) has two scales; one of them is numeric, another – textual. Each scale of a bar chart and its main plotting area can be linked with any number of arbitrary positioned (moveable and rotatable) comments.

Pie chart (`PieChart` class) and a set of rings (`RingSet` class) are similar in design ideas. Each sector has its own comment; this comment can be moved and rotated individually, but when the full complex object is resized, moved forward or rotated, then each sector's comment tries to preserve its relative position to its "parent" sector. Other comments can be linked with an entire pie chart or a set of rings; those comments do not move, when an object is rotated, but they preserve their relative position to the object, when it is moved forward or resized.

The areas of knowledge, which use such plotting, deal with a huge amount of data; the results of analysis depend on the abilities of those, who work with this data. I hear from time to time that the visualization of big amounts of data is not so important, as a sophisticated program can find any tendency by analyzing the data. First, it's nonsense even from the point of analysis; the program can do only what it was ordered to do. The understanding of trends in the huge amount of data is much more efficient, when the computers and the good algorithms for analysis are used, but somewhere next to the computer a small amount of brains is needed and it is a crucial element for the whole analysis. Second, there are a lot of tasks, in which visualization is the main thing.

The plots, which are demonstrated in the **Form_PlotsVariety.cs** are linked with the arrays of data, which were calculated before the opening of the form and are not going to be changed, while you work with it. All the plots are moveable and resizable; there is a huge amount of parameters for visualization, which can be changed by any user at any moment, but the data is not going to change. However, it's not an unbreakable rule. In the same application you can see similar classes, which allow to change in the same easy way (with a mouse) the values, associated with the plots.

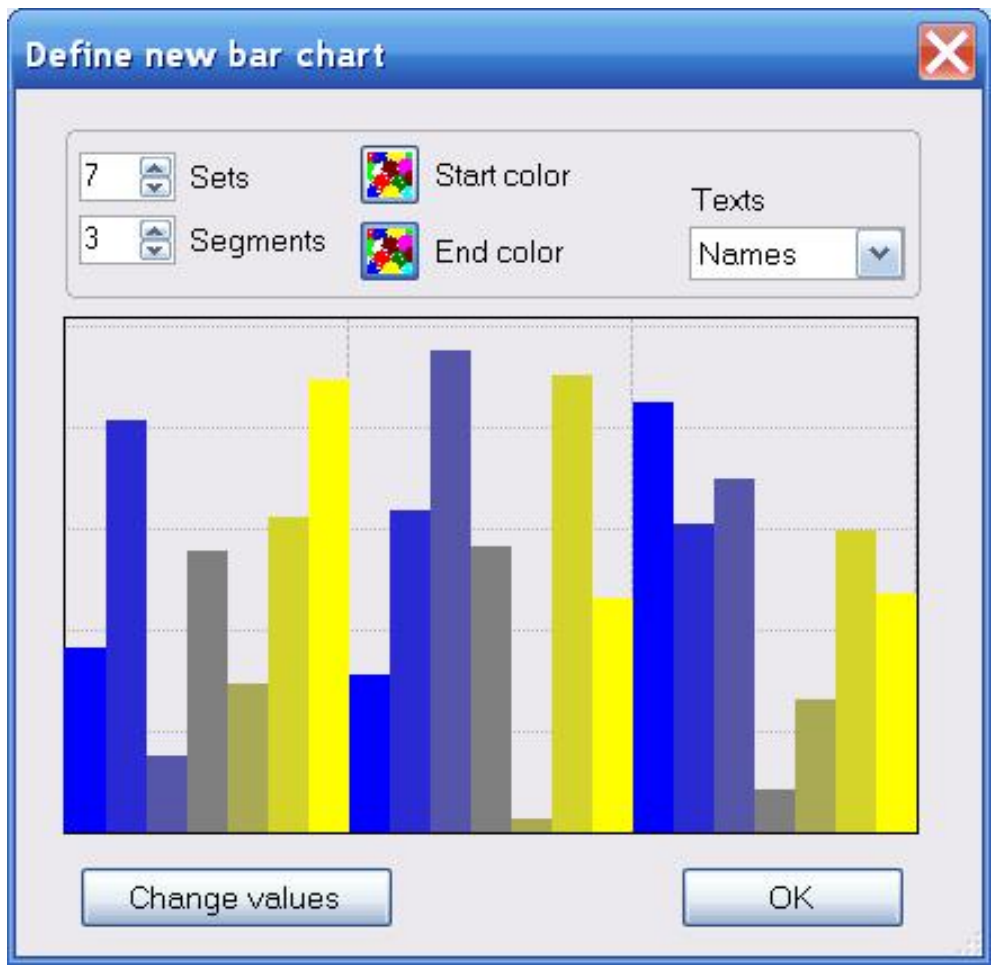

**Fig.9 Form_DefineNewBarChart.cs**

The tuning of different classes, used in the **Form_PlotsVariety.cs**, can be started via different context menus. There are 11 different menus in this form; one of them can be opened at any non-occupied place. Three different positions in this menu allow to open auxiliary forms to define the new bar charts, pie charts or rings.

At **figure 9** you can see the **Form_DefineNewBarChart.cs**, in which the new bar chart can be defined. This form consists of one `ElasticGroup` object, one `PrimitiveBarChart` object, and two moveable / resizable buttons; you can easily transform the form's view to whatever you want.

The bar chart, used here, is called *primitive*, because it has no scales or comments; it can be only moved and resized. But it has one additional feature: its strips can be resized with a mouse, thus changing the values, associated with this bar chart. The strips belong to the `StripInRectangle` class; the cover for such object is simple and consists of two rectangles.

```
public override void DefineCover ()
{
    CoverNode [] nodes = new CoverNode [2];
    switch (linedir)
    {
        case LineDir .Hor:
            float cy = ptAnchor .Y + dh;
            nodes [0] = new CoverNode (0, new RectangleF (cxL, cy - halfsense,
                                           cxR - cxL, 2 * halfsense), Cursors.SizeNS);
            break;
        case LineDir .Ver:
            float cx = ptAnchor .X + dw;
            nodes [0] = new CoverNode (0, new RectangleF (cx - halfsense, cyT,
                                           2 * halfsense, cyB - cyT), Cursors.SizeWE);
            break;
    }
    nodes [1] = new CoverNode (1, new RectangleF (cxL, cyT, cxR - cxL,
                   Math .Max (2, cyB - cyT)), MovementFreedom .None, Cursors.Default);
    cover = new Cover (nodes);
}
```

With such a simple design, we receive an extremely flexible form for developing the new bar charts. I often think that if you have the right approach to the problem, then the code even for the most complex objects would be simple. It's not a mandatory rule, but it often works this way.

**Figure 10** shows similar **Form_DefineNewRing.cs**, which is used to define the new ring; the most interesting object in this form is the ring of the `PrimitiveRing` class. Graphical objects are usually resized by their borders; when the border consists not of the straight lines, but is curved, then the use of the *N-node cover* is the best solution; this technique was already explained in the first article. The cover for the `PrimitiveRing` class belongs to this type and is very similar to the `RingRsRt` class (from the first article), but with an important addition: the borders between the neighbouring sectors are covered by the strip nodes. Thus the borders become moveable and allow to change the sectors (and associated values!) simply with a mouse. It would be better to have the explanation of this moveable border in the first article, but I didn't demonstrate such a sample in the first Demo application, so I'll include some explanation here.

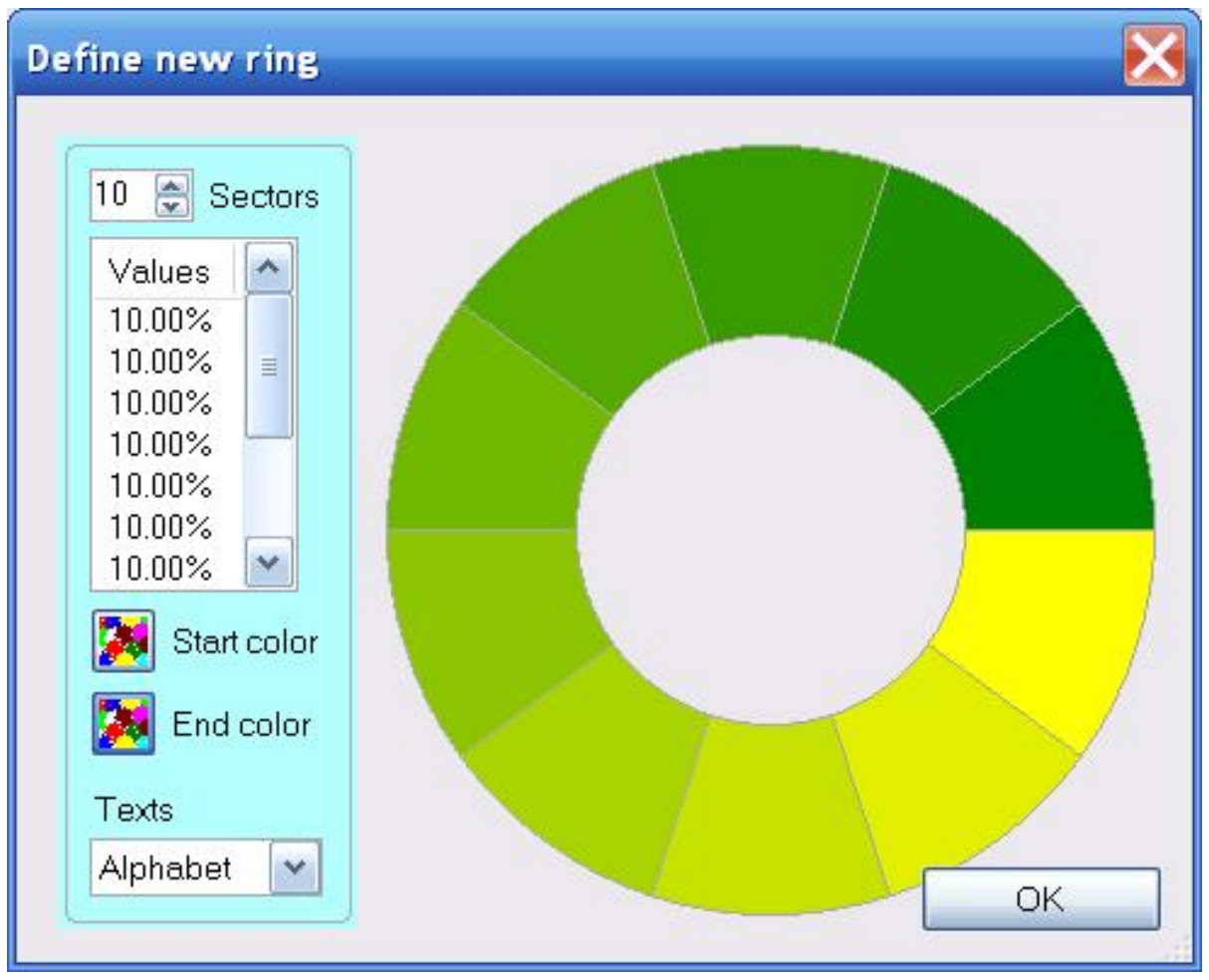

**Fig.10 Form_DefineNewRing.cs**

While writing about the rotation of objects, I showed that at the first moment the `StartRotation()` method must be called. This method receives the coordinates of the point, at which the mouse grabbed an object for rotation. The *compensation* angle is calculated between this point and the object's angle; this compensation doesn't change throughout the object's rotation and allows to redraw the object correctly, while the mouse is going around. Similar technique is used for moving the sectors' borders; at the moment, when the border between two sectors is grabbed for moving, the `StartResectoring()` method of this object must be called to calculate several variables, which are used throughout this moving.

```
private void OnMouseDown (object sender, MouseEventArgs e)
{
    ptMouse_Down = e .Location;
    if (mover .Catch (e .Location, e .Button, bShowAngle))
    {
        GraphicalObject grobj = mover .CaughtSource;
        if (grobj is PrimitiveRing) {
            if (e .Button == MouseButtons .Left) {
                if (mover .CaughtNodeShape == NodeShape .Strip) {
                    ring .StartResectoring (mover .CaughtNode);
                }
            }
            else if (e .Button == MouseButtons .Right) {
                ring .StartRotation (e .Location);
            }
```

When the left button is pressed on the `PrimitiveRing` object and catches the strip node on the border of two sectors, then the node's number is passed to the `StartResectoring()` method. From this parameter, the numbers of the two neighbouring sectors are calculated and two angles, which set the range of moving for the caught border.

If you design a resizable object, there can be a problem with reducing its sizes to total disappearance. The cover of an object depends on its area and the cover's area often is very close to the object's area (in some cases they are identical). If an object totally disappears, then there is some problem with cover definition. Even if you leave some cover for an object, which already disappeared, users don't see anything and don't know about such object. If you really want to delete an object by reducing its size to zero, then you'll also have to exclude such object (its cover) from the mover's queue. Instead I always prefer to set the minimum allowed size of an object and avoid its disappearance in such a way. If an object has some limitations on its movement / resizing, then they are used in the `MoveNode()` method. While resectoring a `PrimitiveRing` object with a mouse, I do not allow the disappearance of any sector and put the minimum allowed angle of 0.05 radian on sectors' size. The checking of the resectoring possibility can be seen in the `PrimitiveRing.MoveNode()` method

```
if (min_angle_Resectoring + 0.05 < angleMouse &&
                                   angleMouse < max_angle_Resectoring - 0.05) {
```

In the previous chapter I mentioned the cases of different moveability for group elements, scales, and comments. In the code of the `PrimitiveRing` and `PrmitivePieChart` classes, you can also see that their covers depend on the declared moveability. The current versions of these classes are designed in such a way, as to allow resectoring, only when an object is declared resizable. In case of need, it's very easy to change slightly the covers for these classes and organize more variants of resizing, for example, by allowing the resizing of objects, but with fixed borders between the sectors.

Objects in the **Form_PlotsVariety.cs** demonstrate different types of plotting and have some interesting features, the programming questions of turning such plots into moveable / resizable (N-node covers, rotation, …) were discussed in the first article. The complex objects, consisting of many parts, must include some kind of instrument to inform the related parts about their movement; then the related parts can adjust their position. The positioning of the parts is unique for each particular class, but the system to inform each other is nearly identical.

While demonstrating the samples of user-driven applications in one area or another, I do not try to show everything that they allow to achieve, but instead I try to show that there are absolutely new horizons of programming, when you base it on moveable / resizable elements. There are millions of good programmers around the world; if their imagination would be not limited any more by the world of fixed design, we will receive such applications, about which up till now we couldn't even think.

# Just for fun

Let's turn to something absolutely different from all the previous samples. There will be no standard plots, no complex functions, no science. Just a small exercise for fun.

I wrote at the beginning that there were no systems or applications totally consisting of the moveable / resizable elements. But were there any programs that came very close to what I call user-driven applications? Yes, there were some, because their purpose was to design an Instrument. And if you develop a good programming instrument, then its behaviour has to be close to the behaviour of all the user-driven applications. A good sample of such type is the **Paint** program. This program provides you with a set of tools (draw a line or curve, fill an area, …). With these tools you can do, whatever you want, and draw anything that your imagination tells you. But *Paint* is an exception among all other programs, because it was purposely thought out and developed to high extent like an instrument. To high extent, but not totally. The *Paint* program provides you with a set of tools, but from the point of the overall construction it is still a designer-driven application.

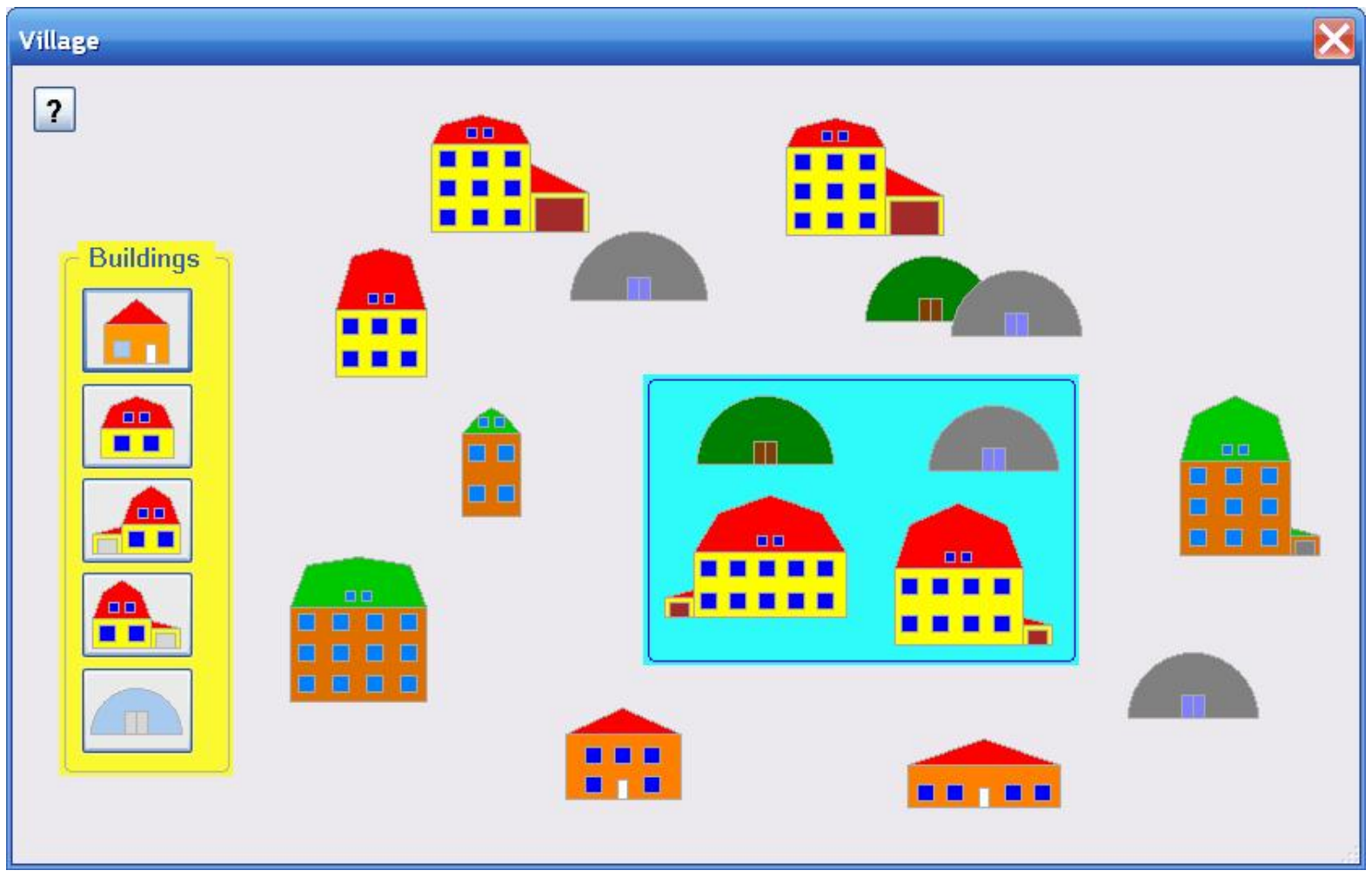

**Fig.11** Your village

The **Form_Village.cs** (**figure 11**) in my demo application works according with all the declared rules of user-driven applications. It is some kind of a painting program, in which you can draw a village. The sketches of the available buildings are included into a group (marked in yellow at **figure 11**); by clicking any of those buttons you add a building of such type to your picture. (Construction of a typical house is cheaper and easier, than building a unique one.) This group is an object of the `ElasticGroup` class, which was used in all the previous samples (with a special exception made in the first sample).

There is another group on the same figure; this group is highlighted in cyan (class `BuildingsGroup`). Any group of buildings can be united into a rectangular group, drawn with a mouse. This group can be moved around and is also useful for several operations, mentioned in its context menu, but I want to underline another feature.

In the first article (in the **Form_Rotation.cs**) I already demonstrated a solution for uniting several screen elements into a group of objects, moving synchronously. In that case the objects, caught inside a drawn rectangle, also move as a group, but to start moving you have to use only the frame. This group from **figure 11** (colored in cyan) can be moved by any inner point, as all the graphical objects are usually moved. There can be different solutions, different implementations, and it's up to the designer to decide about the one, which he wants to use. The decision about the type of moving always influences the development of cover; the latter in its turn influences the drawing procedure. The sample from the first article (class `SimpleFrame`; a group, moved only by frame) has a cover with a big transparent node, slightly less than a frame. The transparent nodes were especially designed to avoid being grabbed by mover; thanks to that node that group can't be moved by the inner points. A `BuildingsGroup` object (**figure 11**, group in cyan) is supposed to be moved by any inner point, so its cover is very simple and consists of one big nonresizable rectangle. But to inform users that this is an object that can be moved by any point, I added the coloring of this area.

# Conclusion

User-driven applications are not some improvement of the currently used programs, but are their alternative. It is another designing philosophy; the step from currently used applications to the user-driven applications is bigger (for USERS!) than from DOS to Windows (or similar systems). The switch to multi-window operational systems, which happened 25 years ago, significantly increased the flexibility of our work with computers, but didn't change the programmer – users relations: users continue to do only whatever was written for them in the designer's scenario. User-driven applications have to pass the main role to the users.

My analysis of the long stagnation in programs' development showed that the problem was in the main design idea. Much more effective applications can be designed only if users get the full control of all the involved screen elements; thus users have to receive the full control of applications.

To give users the full control over all the screen elements, these elements have to become moveable / resizable by users. After an easy to use algorithm of making ANY object moveable and resizable was thought out and applied to the plotting in the scientific / engineering programs, it began to influence the whole design of applications and eventually (with iron logic that didn't allow any steps back or aside) turned them into what is called now the ***user-driven applications***.

Moveability of the elements is not some kind of an extra feature that you simply add to the well-known objects. If you try to design the new program as a standard fixed application and then start adding moveability to the objects one by one, then it would be an absolutely wrong way of development. It will eventually bring you to the right result, but with a huge delay. Moveability of the elements changes the whole system of relations between the objects of your application. Moveability is the main, the fundamental feature of the whole design. You don't add the base of the building after the two floors are already finished; it's better to begin with the basement and put the whole building on top of it.

The design of programs from different areas on the basis of moveable / resizable elements change them in exactly the same way, as it changed the scientific programs. This demonstrates that the rules of these new programs are not special for the area of scientific / engineering programs, but are the common rules for all the user-driven applications.

**Rule 1**. All the elements are moveable.

**Rule 2**. All the viewing parameters must be easily controlled by the users.

**Rule 3**. The users' commands on moving / resizing of objects or on changing the viewing parameters must be implemented exactly as they are; no additions or expanded interpretation by developer are allowed.

**Rule 4**. All the parameters must be saved and restored.

By applying these rules from the start of design, you will get the real user-driven application in the shortest way. The user-driven applications differ from the currently used programs not by some tiny or unimportant details, but by the main ideas:

- Designer has to provide an instrument to solve users' problems. This instrument consists of a correctly working engine (calculation part) and a set of tools to put data inside, to get the results, and to visualize them.
- Designer has to provide a very flexible system of visualization.
- Designer has no right to decide for user, how this instrument has to be used.
- Users get an instrument for solving their tasks and all the possibilities of rearranging the view of this instrument in any way they want.

I have demonstrated the samples of user-driven applications from different areas. There is no single rule for all the programs in the world (except that they have to work), so not all the programs must be turned into user-driven. But a huge amount of applications from different areas have to be turned into user-driven for the benefit of their users.

I think that **rule 3** would be the most controversial one and would be the most criticized by programmers. Simply because it's inside the programmers' heads for decades to prevent any damage to the programs that can be caused by the users' actions. The fool-proof programming is one of the axioms of our work; it's really hard to doubt one of the axioms. In reality, I do not argue with this axiom, I only want it to be used exactly in the range, for which it must be an axiom. Originally it was an axiom for applications' behaviour, which meant calculations; for this area it is absolutely correct. The spreading of the same axiom on the interface (that is one of the origins of dynamic layout) was definitely a mistake. So, continue to develop the fool-proof programs from the purpose of any of them, but don't implement fool-proof ideas, as YOU understand them, into the interface design. Let the programs work absolutely correctly, but be user-driven. An attempt to design user-driven applications with some kind of developers' control of the users' changes in interface would be a mistake. (Some out-of-date doctors continue to insist that it's impossible to be pregnant partly.)

Users don't need to learn, how to deal with the user-driven applications. They have to know only one main rule: EVERYTHING is moveable and tuneable. From there they can work exactly as before.

I am convinced that after the introduction of user-driven applications to the significant amount of users there would be no way back to the fixed applications. It may look strange and unusual at the beginning, but will become normal very quickly. Exactly like it happened with transition from DOS to multi windows systems.

The decision about making a step from the designer-driven to user-driven applications must be made personally by each developer. Each programmer is more a user of other applications than a developer without any connections with the outer world. If the majority of programmers will decide to develop the user-driven applications, we'll find ourselves in another programming world, which gives us – USERS – another level of possibilities.

Dr. Sergey Andreyev ( andreyev_sergey@yahoo.com )

April 2010

# Programs and documents

Several programs and documents are designed for better explanation of moveable and resizable graphics and its use. All files are available at www.SourceForge.net in the project **MoveableGraphics** (names are case sensitive there!). The most important files are renewed from time to time (usually every month); others can be older.

| | |
|---|---|
| **TheoryOfMoveableObjects.zip** | An application (the whole project with all the codes in C#) to accompany the article "On the theory of moveable objects". The article is also included into this ZIP file both in DOC and PDF formats. To run an application, only two files are needed: **TheoryOfMoveableObjects.exe** and **MoveGraphLibrary.dll.** |
| **TheoryOfUserDrivenApplications.zip** | An application (the whole project with all the codes in C#) to accompany the current article. The article (current document) is also included into this ZIP file both in DOC and PDF formats. To run an application, only two files are needed: **TheoryOfUserDrivenApplications.exe** and **MoveGraphLibrary.dll.** |
| **Moveable_Resizable_Objects.doc** | The detailed description of the design and use of moveable / resizable objects. The explanation is based on the samples and codes from the **Test_MoveGraphLibrary** project. 97 pages.<br>The article and the related program are a bit old; they will be renewed in the nearest future (a month or two, I think) |
| **Test_MoveGraphLibrary.zip** | Contains the whole project from Visual Studio with a lot of samples and useful code. The source files are written in C#; all the samples in the **Moveable_Resizable_Objects.doc** are from this project. If you want to run this application, then only two files are needed: **Test_MoveGraphLibrary.exe** and **MoveGraphLibrary.dll.** |
| **MoveGraphLibrary.dll** | The library. |
| **MoveGraphLibrary_Classes.doc** | Description of the classes, included into the **MoveGraphLibrary.dll**. 122 pages. |
| **LiveCalculator.zip** | Calculator, which is included into the **Test_MoveGraphLibrary.exe** for explanation, but presented here as a separate program. To use this application, it must be accompanied by the DLL, which is also inside the zip file. |
| **MoveGraphLibrary_Graphics.doc** | Description of plotting, implemented in the **MoveGraphLibrary.dll**; description of all the tuning dialogs, which are also included into this library. |
| **TuneableGraphics.exe** | This application demonstrates the moveable / resizable objects from absolutely different areas. A year ago it was absolutely different application; now a lot of its forms are similar to the forms from the **Test_MoveGraphLibrary** application, but some are different. |
| **TuneableGraphics_Description.doc** | Description of the **TuneableGraphics** application. |